\documentclass[twocolumn, times, tighten, appendixfloats]{aastex631}

\usepackage{longtable}
\usepackage{natbib}
\usepackage{enumitem}
\usepackage{amsmath}
\usepackage{soul}
\usepackage{float}
\usepackage{romannum}

\let\tablenum\relax\usepackage[load-configurations=astronomy]{siunitx}
\DeclareSIUnit\micron{\ensuremath{\mathrm{\micro\meter}}}
\DeclareSIUnit\sig{\ensuremath{\mathrm{\sigma}}}
\DeclareSIUnit\mag{mag}
\DeclareSIUnit\jansky{Jy}
\DeclareSIUnit\deg{deg}
\DeclareSIUnit\yr{yr}
\DeclareSIUnit\dex{dex}
\DeclareSIUnit\Lsun{\ensuremath{\mathrm{L_\sun}}}
\DeclareSIUnit\Msun{\ensuremath{\mathrm{M_\sun}}}

\begin{document}

\title{First Batch of $\boldsymbol{z\approx 11}$--20 Candidate Objects Revealed by the James Webb Space Telescope Early Release Observations on SMACS 0723-73}

\author[0000-0001-7592-7714]{Haojing Yan}
\affiliation{Department of Physics and Astronomy, University of Missouri-Columbia, Columbia, MO 65211, USA}
\email{yanha@missouri.edu}

\author[0000-0003-3270-6844]{Zhiyuan Ma}
\affiliation{Department of Astronomy, University of Massachusetts, Amherst, MA 01003, USA}

\author[0000-0003-4952-3008]{Chenxiaoji Ling} 
\affiliation{Department of Physics and Astronomy, University of Missouri-Columbia, Columbia, MO 65211, USA}

\author[0000-0003-0202-0534]{Cheng Cheng}
\affiliation{Chinese Academy of Sciences South America Center for Astronomy, National Astronomical Observatories, CAS, Beijing 100101, China}

\author[0000-0001-6511-8745]{Jia-Sheng Huang}
\affiliation{Chinese Academy of Sciences South America Center for Astronomy, National Astronomical Observatories, CAS, Beijing 100101, China}

\begin{abstract}

On July 13, 2022, NASA released to the whole world the data obtained by the 
James Webb Space Telescope ({\it JWST}) Early Release Observations (ERO). These
are the first set of science-grade data from this long-awaited facility, 
marking the beginning of a new era in astronomy. In the study of the early 
universe, JWST will allow us to push far beyond $z\approx 11$, the redshift 
boundary previously imposed by the 1.7~$\mu$m red cut-off of the Hubble Space 
Telescope ({\it HST}). In contrast, JWST's NIRCam reaches $\sim$5~$\mu$m. Among
the {\it JWST} ERO targets there is a nearby galaxy cluster SMACS 0723-73, 
which is a massive cluster and has been long recognized as a potential ``cosmic 
telescope'' in amplifying background galaxies. The ERO six-band NIRCam 
observations on this target have covered an additional flanking field not 
boosted by gravitational lensing, which also sees far beyond {\it HST}. Here we
report the result from our search of candidate objects at $z>11$ using these 
ERO data. In total, there are 87 such objects identified by using the standard
``dropout'' technique. These objects are all detected in multiple bands and 
therefore cannot be spurious. For most of them, their multi-band colors are
inconsistent with known types of contaminants.
If the detected dropout signature is interpreted as the expected 
Lyman-break, it implies that these objects are at $z\approx 11$--20. The large
number of such candidate objects at such high redshifts is not expected from 
the previously favored predictions and demands further investigations.
{\it JWST}\ spectroscopy on such objects will be critical.

\end{abstract}


\section{Introduction}

  Over the past decade, the search of high-redshift (high-$z$) galaxies using
deep {\it HST} images in fields on both ``blank sky'' and lensing clusters
seems to suggest that the number density of galaxies at $z>6$ sharply decreases
towards higher and higher redshifts
\citep[][]{Yan2012,Oesch2012a, Oesch2012b, Finkelstein2012, Zheng2012, Coe2013, McLure2013, Bradley2014, Bouwens2015, McLeod2016,Ishigaki2018}
and even a dearth of galaxies at $z>10$ \citep[][]{Oesch2018}.
This, if true, would create a severe problem for our understanding of the
cosmic hydrogen reionization. There has been a consensus that the reionization
ended at $z\approx 6.2$ \citep[][]{Fan2006, Planck2016reion}. There is also a
piece of evidence from the detection of 1.420 GHz (21 cm) H I absorption of
the cosmic microwave background that first stars came into being at 
$z\approx 17.2$ (spanning $z=13.7$ to 22.9; \cite{Bowman2018}), which is to say
that the reionization should begin at the same redshift. While this result is 
still under debate, an independent line of evidence also points to a similar
redshift as the epoch of first stars. One example is the gravitationally lensed
galaxy MACS1149-JD1 \citep[][]{Zheng2012}, which is at $z=9.1096$ based on the 
ALMA detection of its [OIII] 88~\micron\ line \citep[][]{Hashimoto2018}. The
very existence of oxygen 
\footnote{See also the report of probable detection of carbon and oxygen lines
in GN-z11 \citep[][]{Jiang2021}.}
means that the galaxy has already been polluted by the 
previous generation of stars. To reproduce its spectral energy distribution 
(SED), especially its Balmer break, these authors found that the onset of the 
galaxy must be $\sim$290~Myr earlier, i.e., at the formation redshift 
$z_{f}\approx 15$. It is therefore conceivable that first stars should be 
formed at an even earlier time (by a few tens of Myr), and so should the 
beginning of the reionization. All this, however, is inconsistent with the 
suggestion that the number of galaxies diminishes at $z>10$ because this would
imply that the reionization could not have happened.

  {\it HST}\ cuts off at around 1.7~$\mu$m at the red end, which limits the
highest possible redshift that it could probe to $z\approx 11$. The task of
exploring higher redshifts is now transferred to {\it JWST}, which just 
demonstrated its superb performance through the revelation of its Early 
Release Observations (ERO) on July 12, 2022 \citep[][]{Pontoppidan2022}.
The most relevant ERO observations 
to high-$z$ studies were the NIRCam imaging on the lensing cluster SMACS 
0723-73, which was one of the 41 clusters previously observed by {\it HST}\
in the Reionization Lensing Cluster Survey \citep[RELICS;][]{Coe2019}. 
This field was observed by the NIRCam instrument
(among others) in six bands, and the data were made available world-wide on 
July 13. These were among the very first set of {\it JWST}\ data suitable for
high-$z$ studies. Taking advantage of this opportunity, we report in this
Letter our initial search of $z>11$ objects in this field, using the 
now-standard ``dropout'' technique to identify the characteristic Lyman-break 
signature in the spectral energy distributions (SEDs) of high-$z$ objects
\citep[e.g.][]{Steidel1995, Yan2004b, Stanway2004, Bouwens2004}.
We briefly describe the data and the photometry in 
Section 2 and 3, respectively. The selection of $z>11$ candidate objects is 
presented in Section 4. We conclude with a discussion of the results in 
Section 5.
All magnitudes quoted are in the AB system. All coordinates are of J2000.0 
Equinox. We adopt the following cosmological parameters: 
$\Omega_M=0.27$, $\Omega_\Lambda=0.73$ and $H_0=71$~km~s$^{-1}$~Mpc$^{-1}$.

\section{JWST ERO NIRCam Data SMACS 0723-73} 

\begin{figure*}[htbp]
    \centering
    \includegraphics[width=\textwidth]{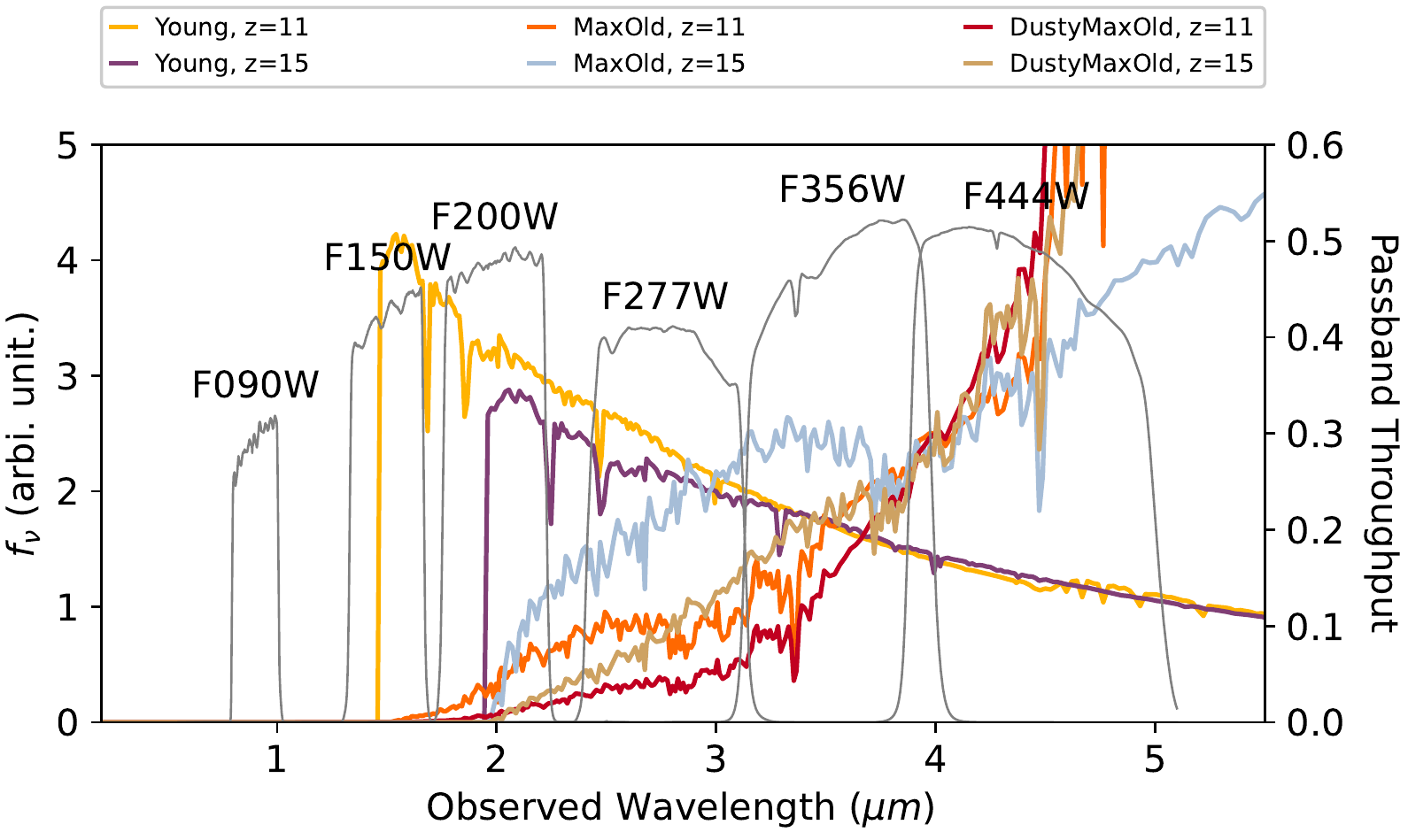}
    \caption{Passbands used in the {\it JWST} ERO NIRCam observations on
    SMACS 0723-73. To demonstrate the dropout technique used for high-$z$ 
    candidate galaxy selection, the spectra of three different model galaxies 
    (as explained in \S 4.1) are superposed. These encompass the bluest and the 
    reddest possibilities. For illustration, they are redshifted to $z=11$ and 
    $z=15$.}
    \label{fig:filt_spec}
\end{figure*}

   The NIRCam observations were carried out on UT June 7, 2022 in six bands,
namely, F090W, F150W and F200W in the ``short wavelength'' (SW) channel and 
F277W, F356W and F444W in the ``long wavelength'' (LW) channel
\footnote{All the {\it JWST} data used in this paper can be found in MAST: \dataset[10.17909/7rjp-th98]{http://dx.doi.org/10.17909/7rjp-th98}.}. 
As NIRCam 
operates in these two channels simultaneously on the same field-of-views
(FOVs), these SW and LW passbands were observed in pairs. The exposures were 
obtained using 9 \texttt{INTRAMODULEBOX} dithers to fill the 
4\arcsec--5\arcsec\ gaps in between the SW detectors, which did not fill the
42\arcsec--48\arcsec\ gap between the modules. This resulted in two
non-overlapping fields of $\sim$2.4\arcmin$\times$2.4\arcmin\ each in SW and 
$\sim$2.2\arcmin$\times$2.2\arcmin\ each in LW. The main part of the cluster
was covered by Module B in the northeastern direction, while Module A
observed a flanking field. The dithered positions
were determined by the \texttt{STANDARD} subpixel dither to optimally sample the
point spread functions (PSFs). For each exposure, the \texttt{MEDIUM8} readout 
pattern was adopted in the ``up-the-ramp'' fitting to determine the count
rate. There was one integration per exposure, and each integration contained
9 groups. In total, the effective exposure time in each band is 7537.2 seconds.
The passband response curves are shown in Figure \ref{fig:filt_spec}, together 
with the spectra of three model galaxies redshifted to $z=11$ and $z=15$, 
which we will explain in \S 4.

   As it turned out, the fully processed ``Stage 3'' data made available 
through the ERO release on July 13 were not suitable for this study due to a
number of reasons. Most critically, the astrometric calibrations of different
bands have small but notable inconsistencies, which makes photometry difficult.
We therefore opted to re-process these data.  We retrieved the
products from the Mikulski Archive for Space Telescopes (MAST), which are the
calibrated single exposures from the standard {\it JWST}\ data reduction 
pipeline after the ``Level 2b'' processing. We then used 
\texttt{calwebb\_image3} (version 1.6.1
\footnote{Three modifications to the pipeline had to be done. The first was to
replace \texttt{photutil} with \texttt{SExtractor} for source detection, the
second was to fix a bug in \texttt{SkyMatch}, and the third was to allow the 
use of an external catalog for astrometric calibration.}), the ``Level 3'' 
pipeline module for imaging data, to combine the single exposures in each band.
As the two modules produced two non-overlapping fields, we processed their data
separately. To obtain internally consistent world coordinate system (WCS), we 
aligned the individual Module B images using the RELICS catalog
\footnote{\url{https://archive.stsci.edu/prepds/relics/\#dataaccess}}
as the reference. For the Module A images, we performed the alignment using the
first image in F150W as the reference. The pixel scale and the image dimensions
were set to 0\arcsec.06~pix$^{-1}$ and 5000$\times$5000 pixels, respectively,
in all bands. The internal alignment among the six mosaics is consistent to
$\lesssim 0.3$~pixel. The absolute astrometric calibration of the final mosaics
in both fields were tied to the second data release of GAIA. The astrometry
thus determined is accurate to $\sim$0\arcsec.03 and 0\arcsec.09 (rms) in 
Module B and A, respectively, in both RA and DEC directions. The accuracy in 
Module A is worse because there is no external source catalog comparable to
the RELICS catalog used as the intermediate in the calibration.

\section{Photometry}

   The final stacks are in the surface brightness units of MJy~sr$^{-1}$. At
the scale of 0\arcsec.06~pix$^{-1}$, this translates to the magnitude zeropoint
of 26.581 for all. We derive that the 5$\sigma$ depths (within 0\arcsec.2 
radius aperture) in F090W, F150W, F200W, F277W, F356W, and F444W are
28.17, 28.54, 28.73, 29.67, 29.73, and 29.72 (28.16, 28.51, 28.69, 29.49, 29.69,
and 29.81)~mag, respectively, for Module A (Module B). The Module B images are
slightly less sensitive than the ones in Module A, presumably due to the
impact of the intra-cluster light (ICL).

   We performed matched aperture photometry by running SExtractor 
\citep{Bertin1996} in the dual-image mode using the F356W images as
the detection images. This choice was made for two reasons. First, the
F356W images are about the deepest (see above). Second, the PSF in
this band is nearly twice as large as in the SW bands, which makes an aperture
defined in this band always sufficient to capture all the source flux in the 
bluer bands. 

  We adopted the \texttt{MAG\_ISO} magnitudes for the color measurements. The
sources of our interest are small enough in the images such that the 
\texttt{MAG\_ISO} apertures include nearly all the source flux while minimizing
the background noise. Hereafter we denote the magnitudes in the six bands as
$m_{090}$, $m_{150}$, $m_{200}$, $m_{277}$, $m_{356}$, and $m_{444}$,
respectively. We only kept the sources that have ${\rm S/N}\geq 5.0$ and
\texttt{ISOAREA\_IMAGE} $\geq 10$ in F356W. 

\section{Selecting high-$z$ candidates as dropouts}

\subsection{Selection Overview}

   Using these data, we searched for candidate high-$z$ objects as F150W,
F200W and F277W dropouts, respectively. If we approximate the throughput 
curves of the NIRCam bands using rectangles, the {\it truncation} of a flat 
spectrum (in $f_\nu$), which is characteristic of Lyman-break at such high
redshifts, will create a color decrement of $\sim$0.75~mag in a blue/red pair
of bands when the break is redshifted out half-way of the blue band (the 
``drop-out'' band). Therefore, we adopted a simple color threshold of
0.8~mag to identify the color decrement caused by Lyman-break, i.e., 
$m_{150} - m_{200} \geq 0.8$, $m_{200} - m_{277} \geq 0.8$ and 
$m_{277} - m_{356} \geq 0.8$~mag for the F150W, F200W and F277W dropouts, 
respectively. To ensure reliable color measurements, objects must have
${\rm S/N\geq 5}$ in the red band next to the break (the ``drop-in'' band),
i.e., in F200W, F277W and F356W for F150W, F200W and F277W dropouts, 
respectively.

   Of course, the size of the decrement in a real galaxy depends on the actual
SED, which can have a wide range of possibilities. This is especially true at
such a high redshift, because the age of the universe is short enough that
activities of short time scales are not averaged out. This is demonstrated in
Figure \ref{fig:filt_spec} by the superposed model galaxy spectra that 
capture the extreme situations. The spectra are generated at $z=11$ and 15, 
respectively, using the population synthesis models of 
\citet[][BC03]{Bruzual2003}. As this is for
demonstration purpose, we only consider their solar metallicity models for
simplicity. There are three models at each redshift. One is a very young (age 
of 10~Myr) galaxy with nearly constant star formation, which represents the 
bluest population that one can get from BC03. We also consider the opposite,
a ``maximally old template'', which is a single burst (``simple stellar 
population'', or SSP) whose age is as old as the age of the universe at the
redshift under discussion, i.e., age of 0.5~Gyr (0.3~Gyr) at $z=11$ ($z=15$).
Such a template has the reddest color among the BC03 models. To make it
even redder, we consider the third template, which is a dusty, maximally old 
template with $A_V=2.0$~mag and reddened according to the extinction law
of \citet[][]{Calzetti2001}. As we will show later, our simple
color criterion tolerates such a wide range of possibilities reasonably well.

   In addition to the 0.8~mag color decrement, we also required
that a valid dropout should be a 2$\sigma$ non-detection in the ``veto'' 
band(s), which is (are) the band(s) to the bluer side of the drop-out band. 
After the initial selections were done, we visually inspected the images of
all candidates in the six bands to reject contaminators due to various
reasons, such as spurious detections around bright objects, image defects
and noise spikes mistakenly included as sources, etc. Due to the nature of
photometric error, some of the reported 2$\sigma$ non-detections in the veto
bands are in fact still visible. Such contaminants were also removed in this
visual inspection step. The surviving dropouts are all detected in at least two
bands, and therefore it is highly unlikely that any of them could be caused by
false detections.

   If we consider the redshift at which the break moves completely
out of the dropout band as the representative redshift of the selection,
our F150W, F200W and F277W dropouts correspond to $z\approx 12.7$
(from 11.3 to 15.4), 17.3 (from 15.4 to 21.8), and 24.7
(from 21.8 to 28.3), respectively. When necessary, we will refer to them
collectively as candidate objects at $z\gtrsim 11$. We have 87 of them in
total, which are presented in the catalog given in Appendix.

    Dropout selections must consider the possible contamination due to two 
types of real objects. One type of them are galaxies at lower redshifts 
dominated by old stellar populations, whose 4000\AA-break could be mistaken as
Lyman-break. In our case here, 4000\AA-break is shifted to 
F150W, F200W and F277W at $z\approx 2.7$, 4.0 and 5.9, respectively. To 
investigate their impact
to our color selections, we use a series of BC03 models redshifted to 
$z=2.6$--6.0 at the step size of 0.1. These models are SSPs and are maximally 
old, e.g., with the age of 2.6~Gyr at $z=2.6$ and 0.9~Gyr at $z=6.0$. We will
refer to them as ``mid-$z$ old galaxies''. The other type of possible 
contaminators are Galactic brown dwarfs, whose strong molecular 
absorption bands could mimic the dropout signature. We use a set of model 
spectra of \citet[][]{Burrows2006}, which cover L and T brown dwarfs with 
effective temperatures ranging from 2300 to 700 K. As we will show later,
the number of possible contaminants caused by either type is very small in our
sample.

\subsection{SED analysis}

    The six-band data afforded us the opportunity to perform SED fitting on
our dropouts. The most important quantities derived from this analysis are the
photometric redshifts ($z_{\rm ph}$), using which we can check whether our
dropouts being at high-$z$ can be supported by an independent method. We note
that this is fitting the SEDs of dropouts that have already been selected,
which is different from selecting high-$z$ candidate galaxies using
$z_{\rm ph}$ directly. Dropout selection is based on identifying the 
Lyman-break signature, which is caused by the Lyman limit and Ly$\alpha$
absorptions due to the intervening H I along the line of sight. Selection based
on $z_{\rm ph}$, on the other hand, depends additionally on the galaxy 
templates in use
(i.e., the assumptions on galaxy properties) as well as the fitting method.
We also note that getting $z_{\rm ph}>11$ for a particular dropout should
not be taken as a confirmation of its being at high-$z$. This is because of the
statistical nature of $z_{\rm ph}$ and that the template set can never be
exhaustive. By the same token, neither should obtaining 
$z_{\rm ph}<11$ for a particular dropout be used as a reason to exclude it 
from the sample. Nevertheless, if we can obtain $z_{\rm ph}>11$ solutions
for the majority of our objects, it will support that they constitute a
legitimate sample at $z>11$.

    To this end, we used {\it Le Phare} \citep[][]{Arnouts1999,Ilbert2006} to 
fit the SEDs of our F150W and F200W dropouts to galaxy templates based on
the BC03 models. This allowed us to derive not only $z_{\rm ph}$ but also
other physical properties of the galaxies, such as age, stellar mass, star
formation rate (SFR), etc. The templates were constructed assuming
exponentially declining star formation histories in 
the form of SFR $\propto e^{-t/\tau}$, where $\tau$ ranges from 0 to 13~Gyr 
(0 for SSP and 13~Gyr to approximate a constant star formation).
These models use the Chabrier initial mass function
\cite[][]{Chabrier2003}.
We adopted the Calzetti extinction law, with $E(B-V)$ ranging from 0 to 1.0
mag. During the fitting, we replaced any ${\rm S/N < 2}$ detections by the
2$\sigma$ upper limits in the relevant bands. For the dropouts in Module B,
we did not correct their SEDs for magnification; such a correction is made
when discussing the statistical properties of stellar mass and SFR.
Overall, this SED analysis shows that most of our objects are consistent
with being at high-$z$. This is judged from the probability distribution 
function (PDF) of $z_{\rm ph}$. The full set of Le Phare fitting results
are presented in Supplementary Materials. 
We will discuss this in some detail in the next two sections.

   Over the past two decade, there have been discussions on the impact of
strong emission lines to broad-band photometric diagnostics
\citep[e.g.,][]{Chary2005, Cardamone2009, Atek2011, Atek2014, Malkan2017,
RB2020}.
This raises the question whether low-$z$ emission-line objects could mimic
high-$z$ dropout colors.
We note that this is unlikely when the dropout search employees enough bands. 
First of all, such a contamination
would require multiple strong emission lines to appear in multiple bands to
elevate the fluxes to the observed level. The only possible combination of
lines are [O II]$\lambda$3727, [O III]$\lambda$5007 and H$\alpha$$\lambda$6563
(and some weaker lines in their vicinities), but the lack of a strong, redder
line makes it difficult to explain dropouts that are detected in four or more
bands. For example, this combination at $z\approx 3.8$--5.0 could elevate the
fluxes in F200W, F277W and F356W to match the brightness of F150W dropouts in
these bands, however the brightness in F444W could not be explained. When a
dropout is detected in only three bands or less (e.g., an F200W or F277W
dropout), such a problem is circumvented; however there are still two other
major difficulties that this contamination scheme can hardly overcome. One is
the null detections in the veto band(s), which means that such a contaminant
would have to be pure emission-line objects without continuum. This implies
that the Ly$\alpha$$\lambda$1216 line should also be very strong. For F200W
dropouts, the only possible redshift range that the [O II]/[O III]/H$\alpha$ 
contamination could occur is $z\approx 5.5$--6.6; but then the Ly$\alpha$ line 
would show up in F090W and the case would be vetoed. Explaining F277W
dropouts in this way has the same difficulty, which leaves
$z\approx 7.4$--9.0 as the only range for this contamination scheme because the 
Ly$\alpha$ line would move to 1.03--1.22~$\mu$m that is in between our 
passbands. However, no such very strong Ly$\alpha$ emitters have been seen
at any redshifts. While there are mechanisms to extinguish Ly$\alpha$ 
line, it remains an open question whether such mechanisms are applicable to an
strong emitter of other lines. The other major difficulty is rareness of strong 
emission-line objects. To alter the broad-band brightness to create the
dropout-like color decrements ($\geq 0.8$~mag), the observed equivalent width
(EW) of a line should be $\geq 3400$\AA\ to $\geq 5150$\AA\ (depending on the
band). This translates to rest-frame EW of $\geq 500$\AA\ (contaminant at
$z\leq 5.8$). At $0.35<z<2.3$, the HST WFC3 Infrared Spectroscopic Parallel
Survey \citep[WISP;][]{Atek2011} has selected objects with strong [OIII] 
or H$\alpha$ lines with rest-frame EW $\geq 200$\AA, and the surface density
is only about 0.2 arcmin$^{-2}$ for those with rest-frame EW $\geq 500$\AA.
Even assuming no redshift evolution, one would get only $\sim$2 such objects 
within our field. This cannot explain the large number of F200W and F277W 
dropouts in our sample. Therefore, we believe that low-$z$ emitters are
not likely a significant source of contamination to any of our three groups of
dropouts. Nevertheless, we still
performed SED fitting using templates including emission lines, which were done
using EAZY \cite[][]{Brammer2008}. 
As expected, the majority of our dropouts also have
high-$z$ solutions.  This set of SED fitting results are also presented in 
Supplementary Materials. In the rest of the main text, we focus on the SED
fitting results using Le Phare with BC03 models.

\subsection{F150W dropouts}

   The depths of the two bands straddling the Lyman-break set the depth of the 
dropout selection. The nominal 2$\sigma$ sensitivities (measured within 
0\arcsec.2 aperture in radius) in F150W are 29.54 and 29.50~mag in Module A and
B, respectively. The criterion of $m_{150}-m_{200}\geq 0.8$~mag for F150W
dropout selection therefore implies that a valid F150W dropout should have 
$m_{200}\leq 28.74$ and 28.70~mag in these two modules, respectively. The
subtlety is that the 0\arcsec.2 aperture is only an approximation of the
real \texttt{MAG\_ISO} apertures in use. Here we ignore this difference.

\begin{figure*}[htbp]
    \centering
    \includegraphics[width=\textwidth]{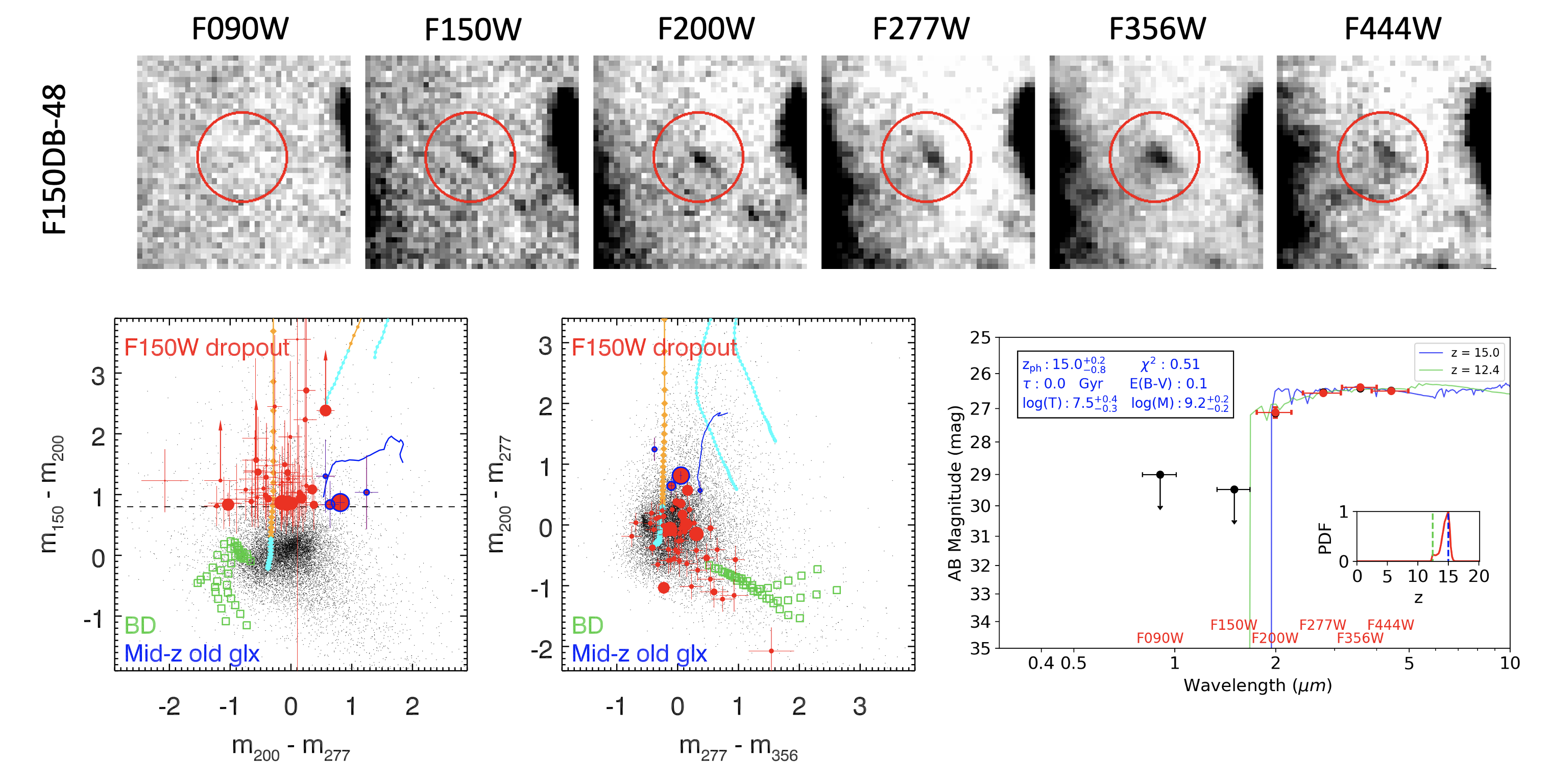}
    \caption{Demonstration of F150W dropouts. The top panel displays the
6-band (as labeled) image cutouts (2\arcsec.4$\times$2\arcsec.4 in size) of 
such an object selected in Module B (its ``short ID'', or ``SID'', is labeled
to left), whose position is indicated by the red circles (0\arcsec.5 in radius).
The SED fitting results of this object is given in the right panel on the
bottom row, where the red symbols are the magnitudes and the downward arrow is
the upper limit. The lower-right inset shows the PDF, which has two peaks
(indicated by the blue and green dashed lines) for this object. The blue and 
green curves are the best-fit template corresponding to the first and second 
peaks ($z_{\rm ph} = 15.0$ and 12.4), respectively. Other best-fit parameters
corresponding to the first peak are given in the upper-left inset. Two 
diagnostic color-color diagrams are given in the left and middle panels on the
bottom row, which show the positions of all the F150W dropouts in these two 
projections of the multi-dimension color space. The filled red circles are the 
dropouts, while the black dots are the field objects. The dashed horizontal
line in the left panel indicates the color threshold of the selection. The
color tracks of the three illustrative templates (as in 
Figure \ref{fig:filt_spec}) are shown from $z=10$ to 30 at the step size of
0.2, and different colors (orange or cyan) are used to indicate redshifts 
above or below $z=11$. The blue curve indicate the region occupied by mid-$z$
old galaxies. The green open boxes are brown dwarfs. In the left panel, four
dropouts (indicated by red circles with blue outline) seem to be close to the
contamination region of mid-$z$ old galaxies. However, in a different 
projection of the color space as shown in the middle panel, three out of these
four are far away from the contamination. In fact, only one dropout (among 60)
is indeed in the contamination region in the color space.
}
    \label{fig:f150d_demo}
\end{figure*}

   There are 32 (28) validated F150W dropouts in Module A (B), and 
the median brightness is $m_{200}=28.23$ (28.13)~mag. There are three and four 
bright objects with $m_{200}\leq 27.5$~mag in Module A and B, respectively
\footnote{To first order, the number of the brightest objects from our result
is not inconsistent with the recently reported results in other 
fields observed by {\it JWST} during the same period \citep[][]{Naidu2022a, 
Castellano2022, Finkelstein2022b}.
}.
Figure \ref{fig:f150d_demo} shows the image stamps of an example object in the
six NIRCam bands (top), together with its SED fitting results (bottom right;
$z_{\rm ph}=15.0$).
Two color-color diagrams are also shown (bottom left and 
middle) to demonstrate the location of these F150W dropouts in the NIRCam color
space. In the $m_{150}-m_{200}$ versus $m_{200}-m_{277}$ projection, which is 
the main diagnostic diagram for F150W dropouts, all the dropouts (filled red
symbols) are far away ($>0.5$~mag) from the contamination region occupied by 
brown dwarfs (open green boxes). Four of them (fill red symbols with blue
outline) seem to be close (within 2-$\sigma$ of color errors) to the mid-$z$
old galaxy contamination region (blue
curve). However, only one of them is really close to this contamination region
in the multi-dimension color space. This is demonstrated in the 
$m_{200}-m_{277}$ versus $m_{277}-m_{356}$ projection of the color space,
where the other three are seen far away from the contamination. This suggests
that the contamination rate is rather low (1 out of 60).

   The SED fitting results also support the high-$z$ interpretation of the
F150W dropouts. 83\% of our objects have the first PDF peak at $z>10$, and
half of the remaining 17\% still have the second PDF peak at $z>10$. Taking
these peak $z_{\rm ph}$ as the solutions, the high-$z$ constituents among the
F150W dropouts have $z_{\rm ph}=10.2$--16.0, with the median of
$z_{\rm ph}=11.6$. The sample has median $m_{200}=28.2$~mag, which corresponds
to rest frame UV absolute magnitude of $M_{UV}=-19.6$~mag.

\begin{figure*}[htbp]
    \centering
    \includegraphics[width=\textwidth]{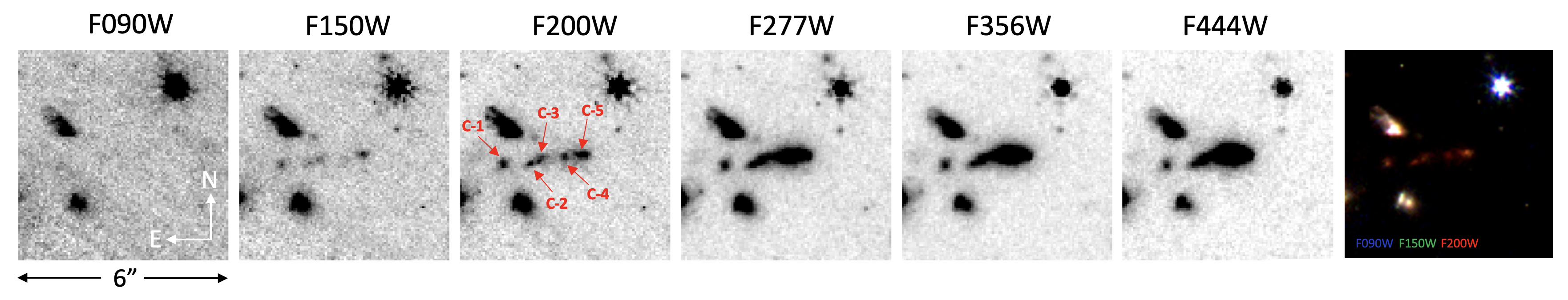}
    \caption{Image stamps of the ``Chain of Five''. The passbands are labeled
    on top. The last panel is a color composite constructed using only the
    short-wavelength bands to preserve the resolution, where the blue, green
    and red colors are F090W, F150W and F200W, respectively. These five objects
    are best seen in F200W, where they are labeled. C-1 is isolated, while the
    other four are severely blended.
    }
    \label{fig:chainof5}
\end{figure*}

   Among the F150W dropouts, one object in the Module B sample is of particular
interest
\footnote{This object is among the ``HST H-dark'' objects in 
\citet[][their ``ES-025'']{Sun2021},
which these authors interpreted as high-mass galaxies at $z\approx 3$--5 
based on their brightness in the {\it Spitzer}\ IRAC images.}.
Due to its complexity, we did not include it in the SED analysis.
Its image is shown in Figure \ref{fig:chainof5}. 
While its structure in the LW
images is not clearly resolved, this object can be discerned as a system of four
separate sources. Together with a close neighbor of similar color, the five
objects line up a ``chain'' in the sky, which we dub as ``C-1'' through 
``C-5'' from East to West.
To obtain better photometry, we extracted these sources using the F200W image
as the detection image, as it shows all five components more clearly owing
to its better resolution. Despite still suffering from severe blending,
this set of photometry suggests that three components, C-2, 3 and 5, are all
consistent with being F150W dropouts. C-4 actually qualifies as a F200W
dropout. C-1, which is barely visible in F090W, also has a rather red color of
$m_{150}-m_{200}=0.67\pm0.08$ suggesting its being at $z\approx 10$.
Combining C-2 through C-5 into a single object, it has $m_{356}=23.1$~mag
($m_{200}=24.8$~mag),
which makes it the brightest system among the high-$z$ candidates in this
field. It deserves spectroscopic study in the near future to determine its
redshift and to understand the relation between the individual components.

\subsection{F200W dropouts}

\begin{figure*}[htbp]
    \centering
    \includegraphics[width=\textwidth]{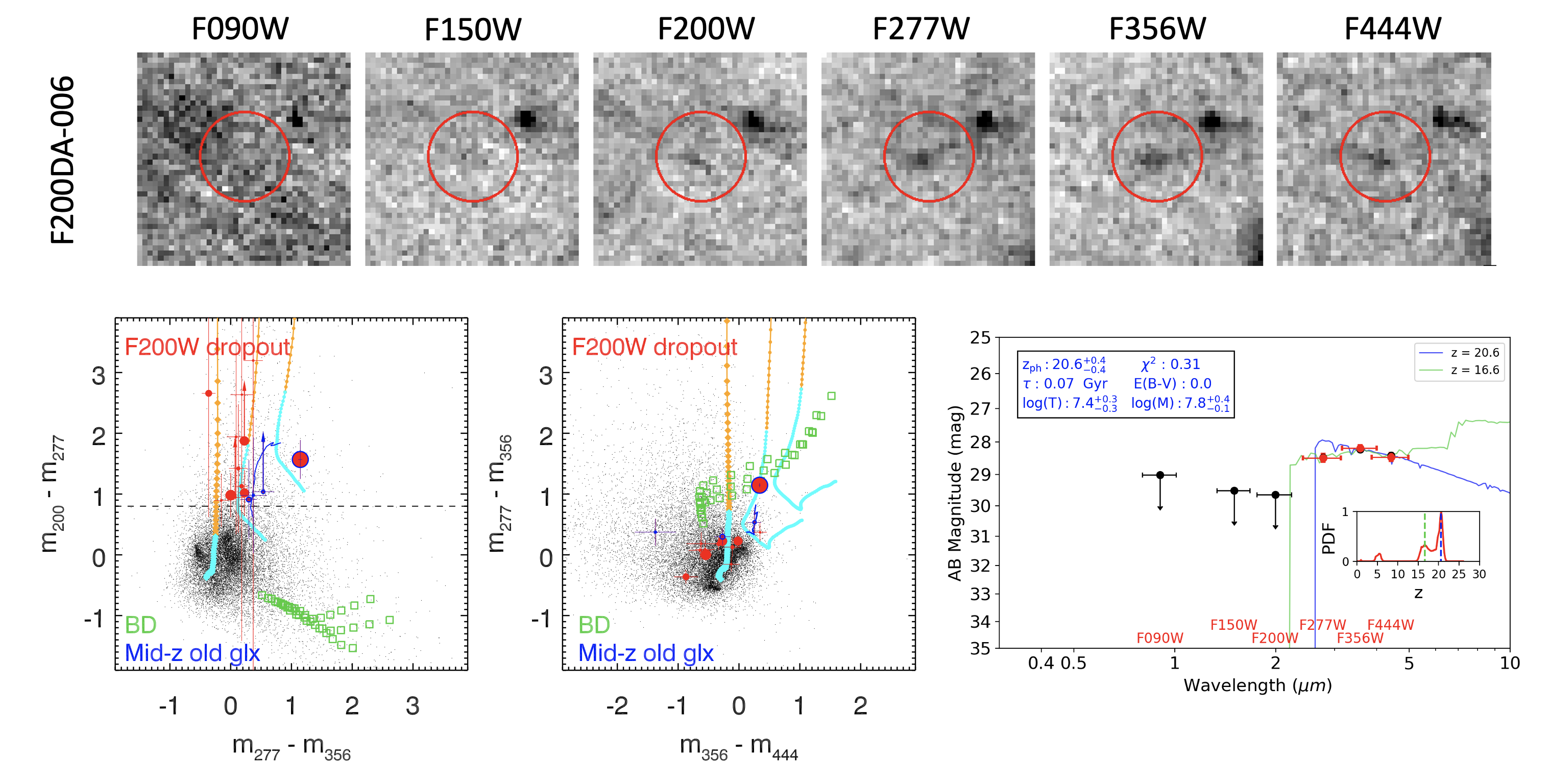}
    \caption{Similar to Figure \ref{fig:f150d_demo}, but for F200W dropouts.
    The example object is from the sample in Module A, and has two PDF peaks
    corresponding to $z_{\rm ph}=20.6$ and 16.6, respectively. In the
    color-color diagrams, the dropout
    symbol sizes are proportional to their brightness in F277W. In the main 
    diagnostic color-color diagram, there are also four dropouts that seem
    to be close to the mid-$z$ old galaxy contamination region, but it can be
    seen from a different projection that most of them actually are not. There
    is only one F200W dropout (among 15 total) that is indeed close to the
    contamination region in the color space.
    }
    \label{fig:f200d_demo}
\end{figure*}

   For F200W dropouts, we consider the 2$\sigma$ sensitivity limits in F200W,
which are 29.73 and 29.69~mag in Module A and B, respectively. Therefore, the 
brightness threshold for a valid F200W dropout is $m_{277}\leq 28.93$ and 
28.89~mag in these two modules, respectively. There are 8 (7) validated F200W 
dropouts in Module A (B). An example object is shown in Figure
\ref{fig:f200d_demo}, together with its SED fitting results ($z_{\rm ph}=20.6$).
Two diagnostic
color-color diagrams are also shown. All these objects are far away from the
brown dwarf contamination region. By coincidence, there are also four objects
close to the mid-$z$ old galaxy contamination region in the main
$m_{200}-m_{277}$ versus $m_{277}-m_{356}$ diagram. However, only one of them
is really close to the contamination region, which can be seen in the
$m_{277}-m_{356}$ versus $m_{356}-m_{444}$ diagram. Therefore, this
suggests that the contamination rate in the F200W dropout sample is also low
(1 out of 15).

   The SED fitting shows that only one F200W dropout has $z_{\rm ph}<10$, and 
the others have $z_{\rm ph}$ ranging from 14.2 to 20.6, with the median of 
$z_{\rm ph}=16.0$. The sample has median $m_{277}=28.6$~mag, which also 
corresponds to $M_{UV}=-19.6$~mag.  There is one bright object from Module A 
that has $m_{277}=26.85\pm0.03$~mag, whose $m_{277}-m_{356}$ color is redder 
than the reddest mid-$z$ old galaxy but is consistent with being a dusty galaxy
at $z>11$. This is the only one from the Module A sample that is brighter than
27.5~mag. Interestingly, no object in the Module B sample is brighter than
$m_{277}=27.5$~mag.

\subsection{F277W dropouts}

   For F277W dropouts, we consider the 2$\sigma$ sensitivity limits in F277W,
which are 30.72 and 30.69~mag in Module A and B, respectively. Therefore, the
brightness threshold for a valid F277W dropout is $m_{356}\leq 29.86$ and 
29.68~mag in these two modules, respectively. There 
are 10 validated objects in the Module A sample but only two in the Module B 
sample. This can be explained by the difference of $\sim$0.2~mag in the 
brightness thresholds of the two modules. Six out of the 10 Module A objects 
are fainter than $m_{356}=29.4$~mag, and therefore it is conceivable that
severe incompleteness at close to the selection limit in Module B 
significantly reduces the number of detections.

\begin{figure*}[htbp]
    \centering
    \includegraphics[width=\textwidth]{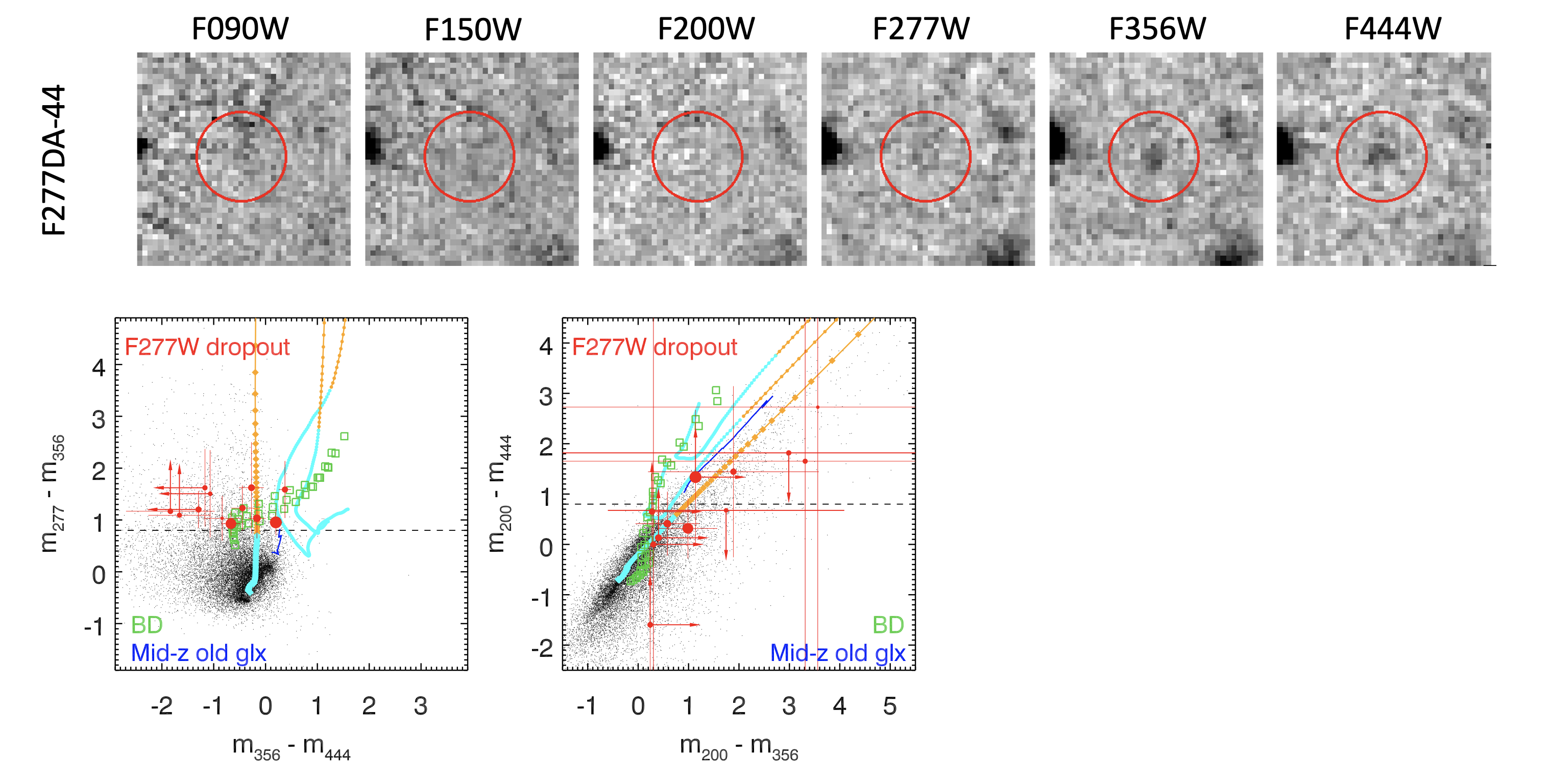}
    \caption{Similar to Figure \ref{fig:f150d_demo}, but for F277W dropouts.
    The example object is selected in Module A. No SED fitting is attempted
    on these dropouts, as they are only detected in two bands. In the
    color-color diagrams, their symbol sizes are proportional to their
    brightness in the F356W band.
    }
    \label{fig:f277d_demo}
\end{figure*}

   As these dropouts have significant detections in only F356W and F444W, we
did not attempt SED fitting. Figure \ref{fig:f277d_demo} shows the image
stamps of one object as example, together with two color-color diagrams. In 
the main diagnostic diagram of $m_{277}-m_{356}$ versus $m_{356}-m_{444}$,
half of these dropouts are far away from the contamination regions and the
other half seem to have similar colors as brown dwarfs. However, these half
are in fact far away from the brown dwarf contamination region in the color
space. This can be seen in the $m_{200}-m_{444}$ versus $m_{200}-m_{356}$
projection. If their observed $m_{277}-m_{356}$ color decrements are indeed due
to Lyman-break, they are at $z\approx 24.7$. 
This suggests that the search for ``first objects'' probably should aim at
$z>20$. Lacking additional information, however, we
refrain from further speculating on their nature.

\section{Discussion}

    For the dropouts selected in Module B, which is on the cluster field, it
is necessary to consider the magnification by the cluster. Shortly before the
ERO release, a lens model for this cluster was published 
\citep[][]{Golubchik2022}. This is based on the \texttt{Light-Traces-Mass} 
approach \citep[LTM;][]{zitrin2009} and uses the RELICS images and
redshifts measured by the Multi Unit Spectroscopic Explorer at the Very Large
Telescope. The ERO NIRCam data have revealed new multiply imaged systems,
which allow a suite of updated lensing models being built using parametric 
modeling codes \citep[][]{Pascale2022, Mahler2022, Caminha2022}. In this work, 
we used the LTM model of \citet[][]{Golubchik2022} and the parametric 
model present in \citep[][]{Pascale2022}. The magnification factor ($\mu$) at a
given source location is the average of the results from these two models
calculated based on $z_{\rm ph}$ as described in \S 4.2.

    It is interesting to compare the dropouts selected in the cluster field
and the flanking field to see whether and how gravitational lensing plays a
role. Overall the cluster field does not produce statistically larger samples;
in fact it has slightly fewer dropouts, which is caused by its slightly 
shallower depths due to the presence of the ICL. Nor does it produce 
statistically brighter samples. However, it does result in the brightest 
candidates: two of its F150W dropouts have $m_{200}<26.5$~mag, one of which
is the ``chain of five'' system. This system has $\mu=2.3$, while the other
has $\mu=1.6$.

   The large number of dropouts presented in this work demands further 
investigations. For simplicity, we concentrate on the F150W and F200W dropouts
in the rest of the discussion. As explained in \S 4, these objects cannot be
fake sources because they are all secure detections in more than two bands
(some are detected in five bands). In the multi-dimension color space, the vast
majority of these sources (with the exception of one object in each sample) are
far away enough from the contamination regions due to mid-$z$ old galaxies and
brown dwarfs, and therefore most of them cannot be attribute to contaminations. 
Recently, \citet[][]{Zavala2022} presented a case where a $z<6$ dusty starburst
mimics the color of a F200W dropout. Their source is detected in millimeter
continuum by NOEMA. In contrast, the deep ALMA 1.3~mm mapping in our field,
which covers $\sim$50\% of Module B, does not reveal any continuum sources at
our dropout positions \citep[see][]{Cheng2022}.

   On the other hand, the color decrements of our dropouts are consistent with 
Lyman-break being shifted to $z\gtrsim 11$. If they are indeed at such high 
redshifts, their large surface densities pose a problem for many predictions.
Over the past decade, it has become a dominant view that there is a paucity of
galaxies at $z>10$. Almost all predictions based on theoretical models or 
extrapolations from the results at lower redshifts expect a rather low surface 
density of galaxies at $z>10$ awaiting {\it JWST} 
\citep[see e.g.][]{Vogelsberger2020, Behroozi2020}. The only exception is
probably the luminosity function proposed by \citet[][]{Yan2010}, which would
predict $\sim$47 galaxies at $11.3\leq z \leq 15.4$ to be discovered as F150W
dropouts over the 5.76~arcmin$^2$ SMACS 0723-73 NIRCam field to the selection
limit of $m_{200}\approx 28.7$~mag as done in this work (but assuming 100\%
completeness). 

   Assuming that the F150W and F200W dropouts are indeed at high-$z$, 
Figure \ref{fig:statistics} (see Appendix) summarizes their properties based on
the SED analysis. The top panel shows the $z_{\rm ph}$ distributions of the 
F150W dropout sample (left) and F200W dropout sample (right). Discarding the
interloppers at $z_{\rm ph}<11$, the median is $z_{\rm ph} =11.6$ and 15.8 for
the F150W and F200W dropouts, respectively, which is largely consistent with
the expectation. The middle and the bottom panels show the distributions of
three derived quantities for the F150W and F200W dropout samples, respectively. 
From left to right, these are age ($T$), stellar mass ($M$) and SFR. The 
objects in the cluster field have been de-magnified to show the intrinsic
$M$ and SFR. The median ages for both samples are $\sim$34 Myr, indicating that
the formation processes of these galaxies had just begun. The median stellar
mass is consistent with this picture, which is $14\times$ and
$8.2\times10^{7}M_\odot$ for the F150W and F200W dropouts, respectively. They
have moderate median SFRs of 19--52~$M_\odot$~yr$^{-1}$, which means that they could
have assembled all their existing stars at such rates over their lifetimes
so far.

   We note that the surface densities of the F150W and F200W dropouts differ
by $4\times$, increasing from 2.6~arcmin$^{-2}$ for the F200W dropouts to
10.6~arcmin$^{-2}$ for the F150W dropouts. This cannot be due to any selection
bias, as the selection of the F200W dropouts in fact has gone $\sim$0.2 mag
deeper (see \S 4.4). If they are at the suggested high redshifts, both samples
probe populations of similar luminosity ($M_{UV} \sim -19.6$~mag).
The SED analysis results shown in
Figure \ref{fig:statistics} also supports that these two samples contain 
similar objects. 
Therefore, assuming that most of these dropouts are genuine $z\gtrsim 11$
galaxies, this would suggest a rapid increase of number density over 
$\sim$~140~Myr from $z\approx 17.3$ to $z\approx 12.7$. 

   We emphasize again that our dropouts are only candidates at $z\gtrsim 11$. 
Even for those that have good SED fits producing $z_{\rm ph}>11$, it is still 
possible that they could be due to some new types of contaminators which we are
not familiar with. It is critical to obtain {\it JWST}\ spectroscopy
on at least a subset of such candidates so that our future exploration of the 
high-$z$ universe can be put onto a solid footing.

%\clearpage

\begin{acknowledgements}

The authors are grateful to the {\it JWST} ERO teams and many behind them for
all their hard work that led to the prompt release of the ERO data used in this
study. We also thank Adi Zitrin for supplying the lensing magnification
calculation.
We thank the anonymous referee for the constructive comments and criticisms,
which improved the quality of this paper.
HY acknowledges the partial support from the University of Missouri
Research Council Grant URC-21-005. ZM is supported in-part by the National
Science Foundation, grant \#1636621. CC is supported by the National Natural
Science Foundation of China, No. 11803044, 12173045. Part of this work is
sponsored by the Chinese Academy of Sciences (CAS), through a grant to the CAS
South America Center for Astronomy (CASSACA).
This project is based on observations made with the NASA/ESA/CSA \emph{James
Webb Space Telescope} and obtained from the Mikulski Archive for Space
Telescopes, which is a collaboration between the Space Telescope Science
Institute (STScI/NASA), the Space Telescope European Coordinating Facility
(ST-ECF/ESA), and the Canadian Astronomy Data Centre (CADC/NRC/CSA).

\end{acknowledgements}

\facility{JWST (NIRCam)}

\appendix

\setcounter{figure}{0}
\renewcommand{\thefigure}{A\arabic{figure}}

\section{Dropout Catalog}

    The full list of dropouts are presented in Table ~\ref{tbl:full_cat}. 
As stated in Section 4.5, the F277W dropouts are only significantly detected in
two to three bands and therefore should be used with caution. Note that the 
Equatorial coordinates are tied to the DR2 of GAIA, which is different
from the astrometric system that the existing RELICS data were calibrated onto.
We also note that the ``chain of five'' system is included in the F150W dropout
sample, however only the C-4 component is listed in the catalog as the 
representative.

\begin{figure*}[htbp]
    \centering
    \includegraphics[width=\textwidth]{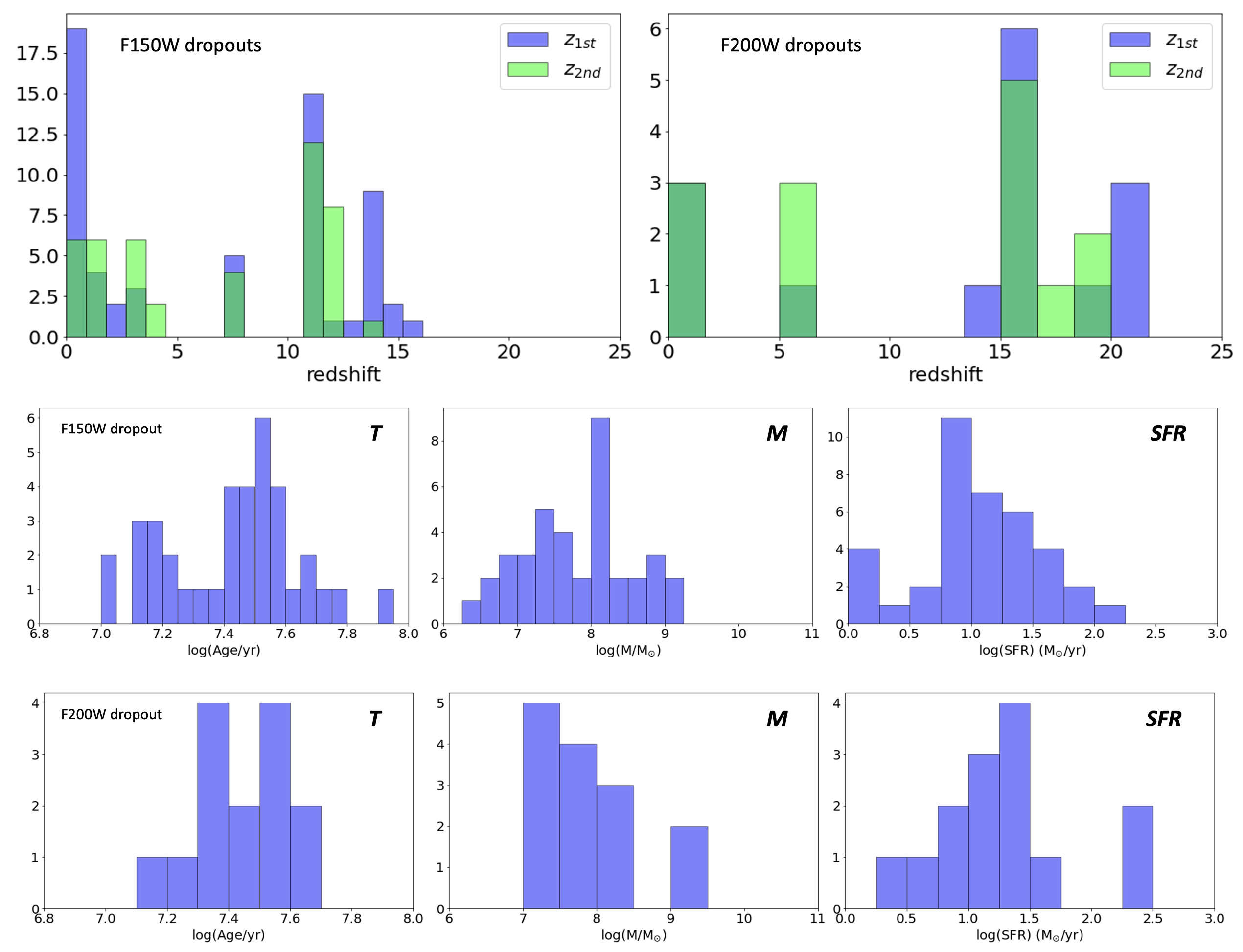}
    \caption{Summary of physical properties of F150W and F200W dropouts as 
    derived by the Le Phare SED analysis. The top panel shows the distribution
    of $z_{\rm ph}$. The best-fit $z_{\rm ph}$, which correspond to the first 
    peaks in the PDFs, are shown in blue. If the secondary peaks exist in the
    PDFs, the corresponding $z_{\rm ph}$ are shown in green.
    The middle and bottom panels show the distributions of age ($T$), stellar 
    mass ($M$) and SFR of the best-fit models corresponding to the first
    PDF peaks. The latter two statistics have taken into
    account the amplification due to lensing in Module B. Stellar masses
    of the $z<11$ interloppers are not included in the middle plot.
    }
    \label{fig:statistics}
\end{figure*}

   The Le Phare SED fitting results of our objects are detailed in Appendix C, 
and the key statistics are summarized in Figure \ref{fig:statistics}. The 
statistics of $T$, $M$ and $SFR$ are done on the subsets that retain only the 
objects that either the primary or the secondary solutions have 
$z_{\rm ph}\geq 11$ (rounded off to integers). If the primary solution has 
$z_{\rm ph}\geq 11$, these parameters are calculated at this solution; if only
the secondary solution has has $z_{\rm ph}\geq 11$, these parameters are
calculated at the secondary solution. In total, the subset contains 39 (14) 
F150W (F200W) dropouts.

   The details of the EAZY SED fitting for $z_{\rm ph}$ are presented in 
Appendix D. Based on the $z_{\rm ph}$ estimates from these two different 
methods, we further group the F150W and F200W dropouts into four categories
(indicated by ``$G_{\rm ph}$'' in Table ~\ref{tbl:full_cat}) based on whether
their $z_{\rm ph}$ are at high-$z$ ($>11.0$ for the F150W dropouts
and $>15.0$ for the F200W dropouts, respectively): I -- both methods have the
primary solution (corresponding to the primary peak of $P(z)$) at high-$z$;
II -- either method has the primary solution at high-$z$; III -- neither
method has the primary solution at high-$z$ but at least one has the secondary
solution (corresponding to the secondary peak of $P(z)$) at high-$z$;
IV -- neither method has the primary nor the secondary solution at high-$z$.
While such a grouping is not necessarily a ranking of the quality of these
dropouts, it could be useful for follow-up studies in the future.

    Three days after our paper was submitted to arXiv, \citet[][]{Atek2022}
and \citet[][]{Donnan2022} also submitted their papers to arXiv and they
discussed high-$z$ objects in SMACS 0723-73 as well. Among the nine
$z\approx 11$--16 candidates in \citet[][]{Atek2022}, their SMACS\_z11b, 
SMACS\_z12b and SMACS\_z11e correspond to our F150DA-63, F150DA-77 and
F150DA-81, respectively. Their SMACS\_z16a and SMACS\_z16b were selected by
us as initial F200W dropouts but were rejected in the visual inspection step
because they are visible in the veto band (F150W). Their SMACS\_z11a,
SMACS\_z11c, SMACS\_z11d, and SMACS\_z12a do not meet our F150W dropout
selection criteria either because of the less prominent color decrement or
due to not having ${\rm S/N\geq 5}$ in F200W. \citet[][]{Donnan2022} also 
included nine $z\approx 11$--16 candidates, but only four of them are in SMACS
0723-73. Their ID 6486 does not meet our F150W dropout color criterion. Their 
21901, 35470, 40079 were selected by us as initial F150W dropouts in Module B
but were later rejected because they all have $m_{200}>28.70$ and are fainter
than the depth threshold that we set for F150W dropout selection in Module B.

    Upon finishing this manuscript, we came to be aware of a possible NIRCam
flux calibration adjustment, which would require re-scaling of the current
pipeline reduced products. In the six bands relevant here, the scaling factors
are 1.143, 1.028, 1.091, 0.870, 0.860, 0.815, from the bluest to the reddest,
respectively. These correspond to adding the following numbers to the
magnitudes derived using the original zeropoint: $-0.145$, $-0.030$, $-0.095$,
0.151, 0.164, and 0.222, from the bluest to the reddest, respectively. As none
of these scaling factors have been vetted and any of them could be evolving in
the next few months, we chose not to make these changes to our photometry. 

    Nonetheless, we still investigate how
such changes could impact our dropout samples. As it turns out, all our F150W
dropouts would still survive, and the new $z_{\rm ph}$ statistics would not
change. Five of the F200W dropouts would have $m_{200}-m_{277}<0.8$~mag
(ranging from 0.65 to 0.78~mag); however, the new $z_{ph}$ estimates would 
still put them at high-$z$. This is understandable because this is equivalent
to lowering the color decrement threshold from 0.8 to 0.65 mag, or in other
words, the Lyman-break signature moves 45\% instead of 50\% out of the
drop-out band. Therefore, none of our conclusions would be changed.

\clearpage
%\movetabledown=2.2in
\begin{longrotatetable}
    \begin{deluxetable*}{cccccccccccc}
      \tablecaption{Catalog of F150W, F200W and F277 Dropouts}
      \label{tbl:full_cat}
        \tabletypesize{\footnotesize}
\tablehead{
    \colhead{ID} & \colhead{Short ID} & \colhead{$m_{090}$} & \colhead{$m_{150}$} & \colhead{$m_{200}$} & \colhead{$m_{277}$} & \colhead{$m_{356}$} & \colhead{$m_{444}$} & \colhead{$\mu$} & \colhead{$G_{\rm ph}$} \\
}
\decimals
\startdata
F150DB J072314.30-732806.78 & F150DB-004 & $>$ 29.15 & $>$ 29.50 & 28.27$\pm$0.18 & 28.28$\pm$0.08 & 28.52$\pm$0.08 & 29.05$\pm$0.12      &  1.92 & I \\
F150DB J072323.97-732758.79 & F150DB-007 & $>$ 29.15 & $>$ 29.50 & 28.08$\pm$0.18 & 27.84$\pm$0.06 & 28.10$\pm$0.06 & 28.47$\pm$0.07      &  2.38 & I \\
F150DB J072305.53-732750.69 & F150DB-013 & $>$ 29.15 & 28.69$\pm$0.28 & 27.76$\pm$0.10 & 28.14$\pm$0.08 & 28.55$\pm$0.10 & 29.47$\pm$0.19 &  2.17 & I \\
F150DB J072312.65-732745.27 & F150DB-021 & $>$ 29.15 & 28.48$\pm$0.34 & 27.39$\pm$0.11 & 27.04$\pm$0.04 & 26.99$\pm$0.03 & 27.26$\pm$0.03 &  3.09 & II \\
F150DB J072305.73-732743.39 & F150DB-023 & $>$ 29.15 & 29.57$\pm$0.54 & 28.65$\pm$0.19 & 29.54$\pm$0.20 & 29.00$\pm$0.10 & 30.24$\pm$0.29 &  2.51 & IV \\
F150DB J072323.74-732740.65 & F150DB-026 & $>$ 29.15 & 29.47$\pm$0.54 & 28.44$\pm$0.18 & 29.08$\pm$0.13 & 28.93$\pm$0.10 & 29.36$\pm$0.13 &  6.38 & II \\
F150DB J072321.44-732736.35 & F150DB-031 & $>$ 29.15 & $>$ 29.50 & 28.67$\pm$0.21 & 28.54$\pm$0.07 & 28.85$\pm$0.08 & 28.90$\pm$0.08      & 14.13 & II \\
F150DB J072330.55-732733.12 & F150DB-033 & $>$ 29.15 & $>$ 29.50 & 28.22$\pm$0.15 & 27.65$\pm$0.04 & 27.28$\pm$0.02 & 27.47$\pm$0.03      &  4.80 & II \\
F150DB J072311.94-732724.97 & F150DB-040 & $>$ 29.15 & 27.87$\pm$0.18 & 27.04$\pm$0.07 & 28.07$\pm$0.08 & 28.30$\pm$0.08 & 28.69$\pm$0.10 & 54.94 & II \\
F150DB J072306.63-732725.45 & F150DB-041 & $>$ 29.15 & $>$ 29.50 & 28.06$\pm$0.19 & 26.82$\pm$0.03 & 27.20$\pm$0.04 & 27.50$\pm$0.05      &  5.18 & I \\
F150DB J072339.32-732722.31 & F150DB-044 & $>$ 29.15 & 29.13$\pm$0.38 & 28.23$\pm$0.14 & 28.21$\pm$0.06 & 28.65$\pm$0.09 & 28.81$\pm$0.08 &  2.19 & II \\
F150DB J072301.57-732718.04 & F150DB-048 & $>$ 29.15 & $>$ 29.50 & 27.12$\pm$0.14 & 26.55$\pm$0.05 & 26.38$\pm$0.03 & 26.50$\pm$0.03      &  3.24 & I \\
F150DB J072324.58-732715.08 & F150DB-050 & $>$ 29.15 & 29.22$\pm$0.47 & 28.36$\pm$0.18 & 28.44$\pm$0.09 & 28.84$\pm$0.10 & 29.69$\pm$0.21 &  7.64 & II \\
F150DB J072328.14-732713.89 & F150DB-052 & $>$ 29.15 & 28.46$\pm$0.39 & 27.63$\pm$0.15 & 27.25$\pm$0.05 & 27.27$\pm$0.04 & 27.81$\pm$0.06 & 16.68 & II \\
F150DB J072312.51-732710.76 & F150DB-054 & $>$ 29.15 & 29.39$\pm$0.39 & 28.55$\pm$0.15 & 28.96$\pm$0.11 & 28.99$\pm$0.09 & 30.13$\pm$0.24 &  9.09 & II \\
F150DB J072307.27-732710.27 & F150DB-056 & $>$ 29.15 & $>$ 29.50 & 28.27$\pm$0.15 & 29.43$\pm$0.22 & 28.50$\pm$0.08 & 28.94$\pm$0.11      &  7.33 & II \\
F150DB J072324.10-732709.84 & F150DB-058 & $>$ 29.15 & 28.32$\pm$0.36 & 27.48$\pm$0.14 & 26.84$\pm$0.04 & 26.94$\pm$0.03 & 27.05$\pm$0.03 & 10.08 & II \\
F150DB J072304.26-732654.24 & F150DB-069 & $>$ 29.15 & $>$ 29.50 & 28.47$\pm$0.15 & 28.87$\pm$0.12 & 29.02$\pm$0.11 & 29.99$\pm$0.25      &  3.10 & I \\
F150DB J072302.23-732641.54 & F150DB-075 & $>$ 29.15 & 27.46$\pm$0.19 & 26.60$\pm$0.08 & 26.67$\pm$0.04 & 26.79$\pm$0.04 & 27.16$\pm$0.05 &  2.23 & I \\
F150DB J072329.42-732639.79 & F150DB-076 & $>$ 29.15 & 29.11$\pm$0.38 & 28.26$\pm$0.15 & 28.42$\pm$0.08 & 28.63$\pm$0.07 & 29.42$\pm$0.13 &  2.16 & II \\
F150DB J072313.16-732629.66 & F150DB-079 & $>$ 29.15 & $>$ 29.50 & 28.07$\pm$0.15 & 28.17$\pm$0.08 & 28.21$\pm$0.07 & 28.55$\pm$0.08      &  2.11 & I \\
F150DB J072322.76-732625.64 & F150DB-082 & $>$ 29.15 & 28.76$\pm$0.31 & 27.88$\pm$0.11 & 27.99$\pm$0.07 & 27.88$\pm$0.05 & 28.17$\pm$0.06 &  1.87 & II \\
F150DB J072307.55-732623.82 & F150DB-084 & $>$ 29.15 & 29.65$\pm$0.48 & 28.49$\pm$0.15 & 28.67$\pm$0.08 & 29.43$\pm$0.15 & 30.51$\pm$0.36 &  1.96 & I \\
F150DB J072314.04-732617.30 & F150DB-088 & $>$ 29.15 & 29.33$\pm$0.27 & 28.19$\pm$0.08 & 27.94$\pm$0.03 & 28.02$\pm$0.03 & 28.10$\pm$0.03 &  1.77 & II \\
F150DB J072326.24-732613.85 & F150DB-090 & $>$ 29.15 & 27.14$\pm$0.18 & 26.27$\pm$0.07 & 25.46$\pm$0.02 & 25.41$\pm$0.01 & 25.42$\pm$0.01 &  1.63 & II \\
F150DB J072324.77-732601.30 & F150DB-095 & $>$ 29.15 & 29.34$\pm$0.45 & 28.47$\pm$0.17 & 28.28$\pm$0.07 & 28.71$\pm$0.09 & 28.84$\pm$0.09 &  1.52 & III \\
F150DB J072325.97-732639.90 & F150DB-C\_4 & $>$ 29.15 & 29.59$\pm$0.30 & 27.65$\pm$0.04 & 25.88$\pm$0.00 & 25.02$\pm$0.00 & 24.36$\pm$0.00 & 2.28 & \nodata \\
F200DB J072307.67-732801.58 & F200DB-015 & $>$ 29.15 & $>$ 29.50 & $>$ 29.69 & 28.74$\pm$0.12 & 28.67$\pm$0.10 & 29.31$\pm$0.16           &  1.94 & I \\
F200DB J072322.77-732739.72 & F200DB-045 & $>$ 29.15 & $>$ 29.50 & $>$ 29.69 & 27.82$\pm$0.08 & 27.59$\pm$0.05 & 27.86$\pm$0.06           &  7.86 & I \\
F200DB J072306.42-732719.88 & F200DB-086 & $>$ 29.15 & $>$ 29.50 & 28.88$\pm$0.37 & 27.86$\pm$0.08 & 27.63$\pm$0.05 & 27.65$\pm$0.05      &  5.43 & II \\
F200DB J072337.04-732712.23 & F200DB-109 & $>$ 29.15 & $>$ 29.50 & $>$ 29.69 & 28.88$\pm$0.18 & 28.51$\pm$0.12 & 29.88$\pm$0.32           &  2.54 & II \\
F200DB J072325.35-732646.05 & F200DB-159 & $>$ 29.15 & $>$ 29.50 & $>$ 29.69 & 28.77$\pm$0.12 & 28.68$\pm$0.09 & 29.24$\pm$0.14           &  2.78 & I \\
F200DB J072311.09-732638.03 & F200DB-175 & $>$ 29.15 & $>$ 29.50 & $>$ 29.69 & 28.18$\pm$0.07 & 28.55$\pm$0.08 & 29.42$\pm$0.18           &  2.71 & I \\
F200DB J072312.62-732631.73 & F200DB-181 & $>$ 29.15 & $>$ 29.50 & 28.63$\pm$0.36 & 27.65$\pm$0.07 & 27.65$\pm$0.06 & 28.19$\pm$0.08      &  2.20 & I \\
F277DB J072317.55-732825.26 & F277DB-001 & $>$ 29.15 & $>$ 29.50 & $>$ 29.69 & 31.00$\pm$0.54 & 29.41$\pm$0.10 & 29.04$\pm$0.07           & \nodata & \nodata \\
F277DB J072308.41-732622.72 & F277DB-013 & $>$ 29.15 & $>$ 29.50 & $>$ 29.69 & 30.24$\pm$0.37 & 29.21$\pm$0.11 & 29.38$\pm$0.13           & \nodata & \nodata \\
F150DA J072241.01-732955.00 & F150DA-005 & $>$ 29.16 & 28.88$\pm$0.31 & 28.05$\pm$0.12 & 28.19$\pm$0.05 & 28.36$\pm$0.06 & 28.28$\pm$0.06 & \nodata & I \\
F150DA J072244.88-732953.69 & F150DA-007 & $>$ 29.16 & $>$ 29.54 & 28.38$\pm$0.16 & 28.52$\pm$0.07 & 28.49$\pm$0.06 & 28.64$\pm$0.08      & \nodata & I \\
F150DA J072252.75-732951.67 & F150DA-008 & $>$ 29.16 & $>$ 29.54 & 27.76$\pm$0.21 & 28.30$\pm$0.11 & 27.83$\pm$0.07 & 28.51$\pm$0.12      & \nodata & II \\
F150DA J072240.09-732946.14 & F150DA-010 & $>$ 29.16 & $>$ 29.54 & 28.49$\pm$0.21 & 28.38$\pm$0.07 & 28.73$\pm$0.10 & 28.31$\pm$0.07      & \nodata & I \\
F150DA J072236.76-732935.68 & F150DA-013 & $>$ 29.16 & 29.30$\pm$0.48 & 28.22$\pm$0.15 & 28.96$\pm$0.12 & 28.64$\pm$0.08 & 28.77$\pm$0.10 & \nodata & II \\
F150DA J072244.74-732926.87 & F150DA-015 & $>$ 29.16 & 29.46$\pm$0.53 & 28.54$\pm$0.19 & 28.51$\pm$0.07 & 28.32$\pm$0.06 & 28.83$\pm$0.09 & \nodata & IV \\
F150DA J072256.03-732921.94 & F150DA-018 & $>$ 29.16 & $>$ 29.54 & 27.97$\pm$0.18 & 28.54$\pm$0.14 & 27.59$\pm$0.05 & 27.93$\pm$0.07      & \nodata & II \\
F150DA J072239.40-732920.50 & F150DA-019 & $>$ 29.16 & 29.40$\pm$0.47 & 28.50$\pm$0.17 & 28.65$\pm$0.08 & 28.70$\pm$0.09 & 29.02$\pm$0.11 & \nodata & III \\
F150DA J072255.88-732917.48 & F150DA-020 & $>$ 29.16 & 29.32$\pm$0.34 & 28.22$\pm$0.10 & 28.67$\pm$0.07 & 28.66$\pm$0.06 & 28.46$\pm$0.06 & \nodata & I \\
F150DA J072233.47-732909.57 & F150DA-024 & $>$ 29.16 & 28.93$\pm$0.30 & 28.12$\pm$0.12 & 29.34$\pm$0.17 & 28.60$\pm$0.08 & 29.02$\pm$0.12 & \nodata & IV \\
F150DA J072246.02-732908.13 & F150DA-026 & $>$ 29.16 & $>$ 29.54 & 28.42$\pm$0.18 & 29.01$\pm$0.13 & 29.16$\pm$0.14 & 28.80$\pm$0.10      & \nodata & I \\
F150DA J072301.03-732907.20 & F150DA-027 & $>$ 29.16 & 29.80$\pm$0.50 & 28.57$\pm$0.14 & 30.64$\pm$0.36 & 29.10$\pm$0.09 & 30.31$\pm$0.27 & \nodata & IV \\
F150DA J072240.65-732900.53 & F150DA-031 & $>$ 29.16 & $>$ 29.54 & 28.23$\pm$0.15 & 28.27$\pm$0.07 & 28.16$\pm$0.06 & 28.02$\pm$0.05      & \nodata & I \\
F150DA J072300.68-732848.43 & F150DA-036 & $>$ 29.16 & 29.25$\pm$0.34 & 28.30$\pm$0.12 & 28.85$\pm$0.10 & 28.91$\pm$0.10 & 28.82$\pm$0.09 & \nodata & I \\
F150DA J072302.96-732846.18 & F150DA-038 & $>$ 29.16 & $>$ 29.54 & 28.64$\pm$0.21 & 29.24$\pm$0.16 & 29.21$\pm$0.15 & 29.36$\pm$0.16      & \nodata & I \\
F150DA J072300.58-732847.04 & F150DA-039 & $>$ 29.16 & 29.00$\pm$0.37 & 28.01$\pm$0.13 & 28.43$\pm$0.09 & 28.13$\pm$0.06 & 28.30$\pm$0.08 & \nodata & II \\
F150DA J072250.08-732851.05 & F150DA-047 & $>$ 29.16 & 29.59$\pm$0.40 & 28.73$\pm$0.16 & 29.83$\pm$0.21 & 29.23$\pm$0.11 & 29.84$\pm$0.20 & \nodata & IV \\
F150DA J072245.00-732836.90 & F150DA-050 & $>$ 29.16 & $>$ 29.54 & 27.94$\pm$0.14 & 27.69$\pm$0.06 & 27.49$\pm$0.05 & 27.45$\pm$0.05      & \nodata & I \\
F150DA J072226.94-732833.82 & F150DA-052 & $>$ 29.16 & 29.59$\pm$0.54 & 28.56$\pm$0.17 & 28.49$\pm$0.09 & 28.41$\pm$0.08 & 28.53$\pm$0.08 & \nodata & II \\
F150DA J072232.48-732833.23 & F150DA-053 & $>$ 29.16 & $>$ 29.54 & 28.61$\pm$0.15 & 29.19$\pm$0.11 & 29.30$\pm$0.12 & 29.43$\pm$0.13      & \nodata & I \\
F150DA J072238.89-732830.88 & F150DA-054 & $>$ 29.16 & 29.52$\pm$0.48 & 28.64$\pm$0.18 & 28.89$\pm$0.09 & 28.83$\pm$0.08 & 29.19$\pm$0.11 & \nodata & III \\
F150DA J072258.72-732828.40 & F150DA-057 & $>$ 29.16 & 27.80$\pm$0.20 & 26.94$\pm$0.08 & 26.93$\pm$0.03 & 26.79$\pm$0.03 & 27.09$\pm$0.03 & \nodata & II \\
F150DA J072248.28-732827.38 & F150DA-058 & $>$ 29.16 & $>$ 29.54 & 27.87$\pm$0.15 & 27.92$\pm$0.09 & 27.38$\pm$0.05 & 27.57$\pm$0.06      & \nodata & I \\
F150DA J072240.76-732823.77 & F150DA-060 & $>$ 29.16 & 29.59$\pm$0.44 & 28.64$\pm$0.16 & 28.69$\pm$0.08 & 28.64$\pm$0.07 & 28.75$\pm$0.08 & \nodata & I \\
F150DA J072254.23-732823.59 & F150DA-062 & $>$ 29.16 & 28.97$\pm$0.44 & 27.94$\pm$0.14 & 27.80$\pm$0.04 & 28.03$\pm$0.05 & 27.72$\pm$0.04 & \nodata & I \\
F150DA J072253.83-732823.24 & F150DA-063 & $>$ 29.16 & 27.50$\pm$0.26 & 26.63$\pm$0.10 & 26.78$\pm$0.04 & 26.47$\pm$0.03 & 26.54$\pm$0.03 & \nodata & II \\
F150DA J072239.62-732812.19 & F150DA-066 & $>$ 29.16 & 28.94$\pm$0.23 & 28.06$\pm$0.09 & 28.71$\pm$0.07 & 29.04$\pm$0.09 & 29.05$\pm$0.09 & \nodata & II \\
F150DA J072238.35-732757.11 & F150DA-075 & $>$ 29.16 & $>$ 29.54 & 28.08$\pm$0.16 & 28.49$\pm$0.11 & 27.84$\pm$0.06 & 27.88$\pm$0.06      & \nodata & II \\
F150DA J072252.23-732755.40 & F150DA-077 & $>$ 29.16 & $>$ 29.54 & 28.23$\pm$0.10 & 28.50$\pm$0.07 & 28.48$\pm$0.06 & 28.58$\pm$0.07      & \nodata & I \\
F150DA J072249.25-732749.89 & F150DA-078 & $>$ 29.16 & 28.87$\pm$0.37 & 28.01$\pm$0.14 & 27.98$\pm$0.06 & 27.71$\pm$0.04 & 28.15$\pm$0.07 & \nodata & II \\
F150DA J072249.24-732744.55 & F150DA-081 & $>$ 29.16 & $>$ 29.54 & 28.48$\pm$0.16 & 28.49$\pm$0.07 & 28.35$\pm$0.06 & 28.44$\pm$0.06      & \nodata & I \\
F150DA J072252.78-732741.93 & F150DA-082 & $>$ 29.16 & 29.11$\pm$0.35 & 28.25$\pm$0.14 & 29.26$\pm$0.16 & 29.04$\pm$0.12 & 28.99$\pm$0.11 & \nodata & III \\
F150DA J072242.72-732732.31 & F150DA-083 & $>$ 29.16 & 28.21$\pm$0.28 & 27.27$\pm$0.10 & 27.10$\pm$0.04 & 27.02$\pm$0.04 & 27.54$\pm$0.06 & \nodata & II \\
F200DA J072240.35-733010.35 & F200DA-006 & $>$ 29.16 & $>$ 29.54 & $>$ 29.73 & 28.48$\pm$0.11 & 28.19$\pm$0.08 & 28.46$\pm$0.11           & \nodata & I \\
F200DA J072243.92-732915.78 & F200DA-033 & $>$ 29.16 & $>$ 29.54 & 28.42$\pm$0.32 & 26.85$\pm$0.03 & 25.70$\pm$0.01 & 25.36$\pm$0.01      & \nodata & IV \\
F200DA J072305.21-732913.40 & F200DA-034 & $>$ 29.16 & $>$ 29.54 & $>$ 29.73 & 28.85$\pm$0.15 & 28.66$\pm$0.11 & 29.28$\pm$0.21           & \nodata & II \\
F200DA J072303.93-732906.15 & F200DA-040 & $>$ 29.16 & $>$ 29.54 & $>$ 29.73 & 28.61$\pm$0.12 & 28.48$\pm$0.10 & 28.59$\pm$0.11           & \nodata & I \\
F200DA J072237.03-732841.58 & F200DA-056 & $>$ 29.16 & $>$ 29.54 & 29.71$\pm$0.43 & 28.81$\pm$0.08 & 28.97$\pm$0.09 & 29.09$\pm$0.10      & \nodata & II \\
F200DA J072231.70-732838.66 & F200DA-061 & $>$ 29.16 & $>$ 29.54 & 29.65$\pm$0.49 & 28.52$\pm$0.08 & 28.34$\pm$0.06 & 28.68$\pm$0.08      & \nodata & III \\
F200DA J072232.43-732806.80 & F200DA-089 & $>$ 29.16 & $>$ 29.54 & $>$ 29.73 & 28.81$\pm$0.13 & 28.44$\pm$0.09 & 28.09$\pm$0.06           & \nodata & I \\
F200DA J072234.80-732800.23 & F200DA-098 & $>$ 29.16 & $>$ 29.54 & $>$ 29.73 & 28.69$\pm$0.16 & 28.16$\pm$0.09 & 27.90$\pm$0.07           & \nodata & I \\
F277DA J072247.81-733004.68 & F277DA-001 & $>$ 29.16 & $>$ 29.54 & $>$ 29.73 & $>$ 30.66 & 29.49$\pm$0.15 & $>$ 30.72                     & \nodata & \nodata \\
F277DA J072225.29-732854.54 & F277DA-025 & $>$ 29.16 & $>$ 29.54 & $>$ 29.73 & $>$ 30.66 & 29.55$\pm$0.16 & $>$ 30.72                     & \nodata & \nodata \\
F277DA J072241.25-732842.96 & F277DA-028 & $>$ 29.16 & $>$ 29.54 & $>$ 29.73 & 30.62$\pm$0.42 & 29.39$\pm$0.12 & 29.84$\pm$0.18           & \nodata & \nodata \\
F277DA J072300.29-732830.55 & F277DA-033 & $>$ 29.16 & $>$ 29.54 & $>$ 29.73 & $>$ 30.66 & 29.65$\pm$0.18 & $>$ 30.72                     & \nodata & \nodata \\
F277DA J072235.39-732821.47 & F277DA-040 & $>$ 29.16 & $>$ 29.54 & $>$ 29.73 & 30.84$\pm$0.41 & 29.81$\pm$0.15 & 30.65$\pm$0.32           & \nodata & \nodata \\
F277DA J072256.53-732811.17 & F277DA-044 & $>$ 29.16 & $>$ 29.54 & $>$ 29.73 & 29.55$\pm$0.24 & 28.59$\pm$0.09 & 28.39$\pm$0.08           & \nodata & \nodata \\
F277DA J072252.57-732807.65 & F277DA-045 & $>$ 29.16 & $>$ 29.54 & $>$ 29.73 & 30.63$\pm$0.34 & 29.43$\pm$0.11 & $>$ 30.72                & \nodata & \nodata \\
F277DA J072233.84-732800.85 & F277DA-046 & $>$ 29.16 & $>$ 29.54 & $>$ 29.73 & $>$ 30.66 & 29.33$\pm$0.15 & 29.60$\pm$0.19                & \nodata & \nodata \\
F277DA J072237.79-732758.63 & F277DA-047 & $>$ 29.16 & $>$ 29.54 & $>$ 29.73 & 29.70$\pm$0.22 & 28.77$\pm$0.09 & 29.44$\pm$0.16           & \nodata & \nodata \\
F277DA J072242.11-732754.78 & F277DA-049 & $>$ 29.16 & $>$ 29.54 & $>$ 29.73 & $>$ 30.66 & 29.57$\pm$0.13 & $>$ 30.72                     & \nodata & \nodata \\
\enddata
\end{deluxetable*}
\end{longrotatetable}

\setcounter{figure}{0}
\renewcommand{\thefigure}{B\arabic{figure}}

\section{Dropout Images}

   Figures \ref{fig:all_f150d_stamps}, \ref{fig:all_f200d_stamps} and
\ref{fig:all_f277d_stamps} present the image cutouts of all our F150W, F200W
and F277W dropouts, respectively. The ``chain of five'' is already presented
in Figure \ref{fig:chainof5} and is not included here.

\begin{figure*}[htbp]
    \centering
    \includegraphics[height=\textheight]{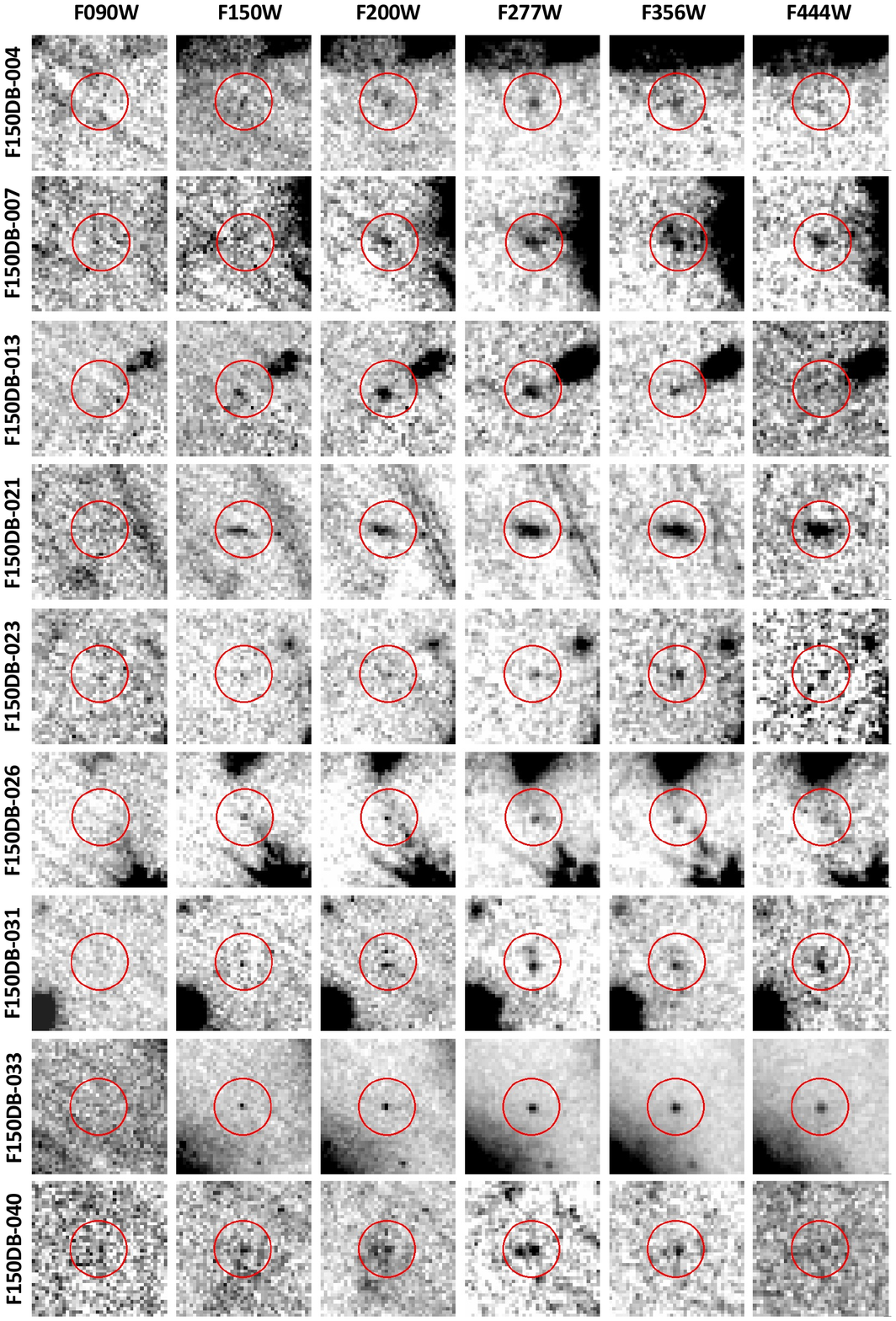}
    \caption{Image cutouts of F150W dropouts in fix NIRCam bands. The images
    are 2\arcsec.4$\times$2\arcsec.4 in size, and the dropout positions are
    indicated by the red circles (0\arcsec.5 in radius). The dropout SIDs
    are labeled to left.
    }
    \label{fig:all_f150d_stamps}
\end{figure*}

\setcounter{figure}{0}
\begin{figure*}[htbp]
%\ContinuedFloat 
    \centering
    \includegraphics[height=\textheight]{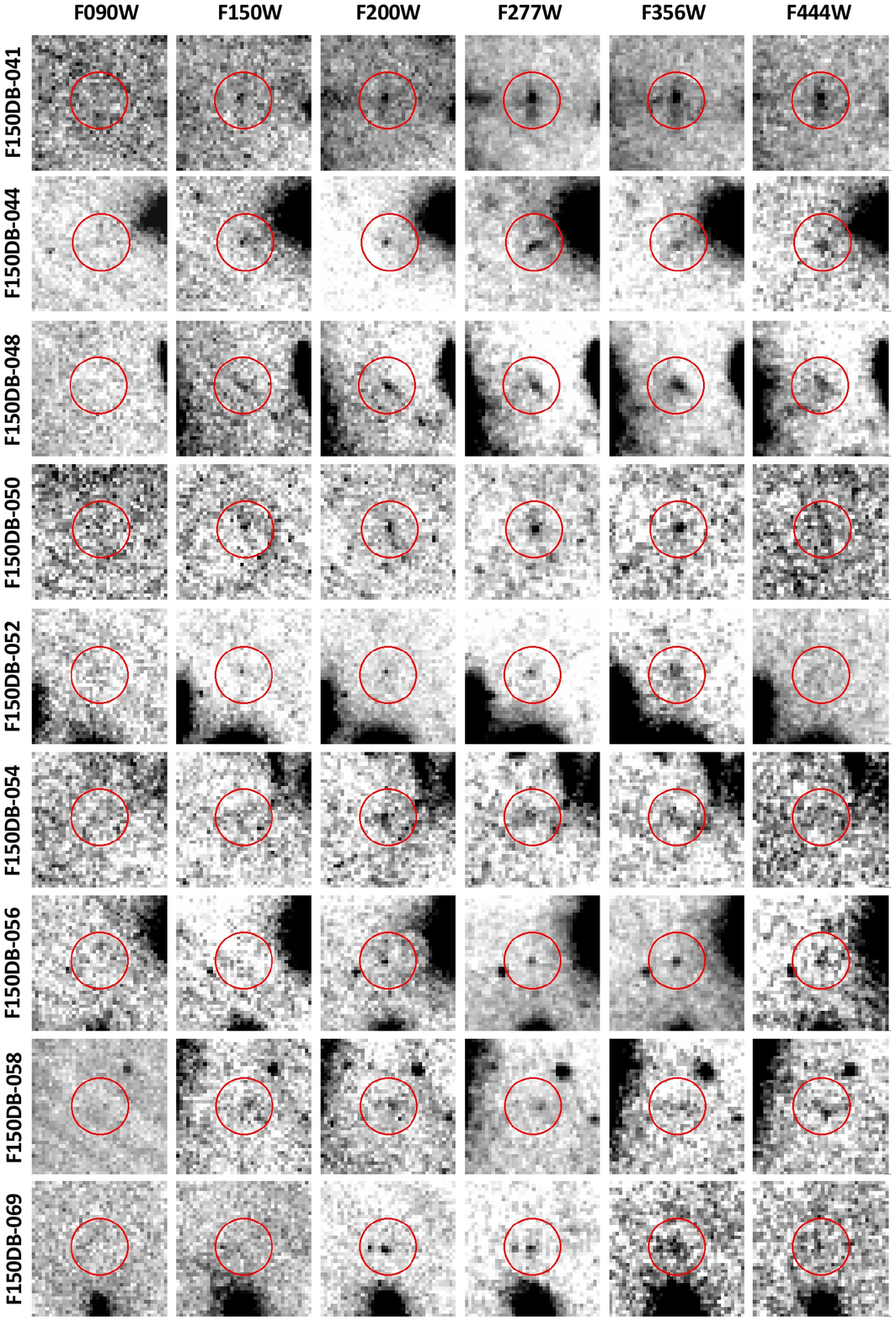}
    \caption{(cont.) 
    }
  %\label{fig:all_f150d_stamps}
\end{figure*}

\setcounter{figure}{0}
\begin{figure*}[htbp]
%\ContinuedFloat 
    %\centering
    \includegraphics[height=\textheight]{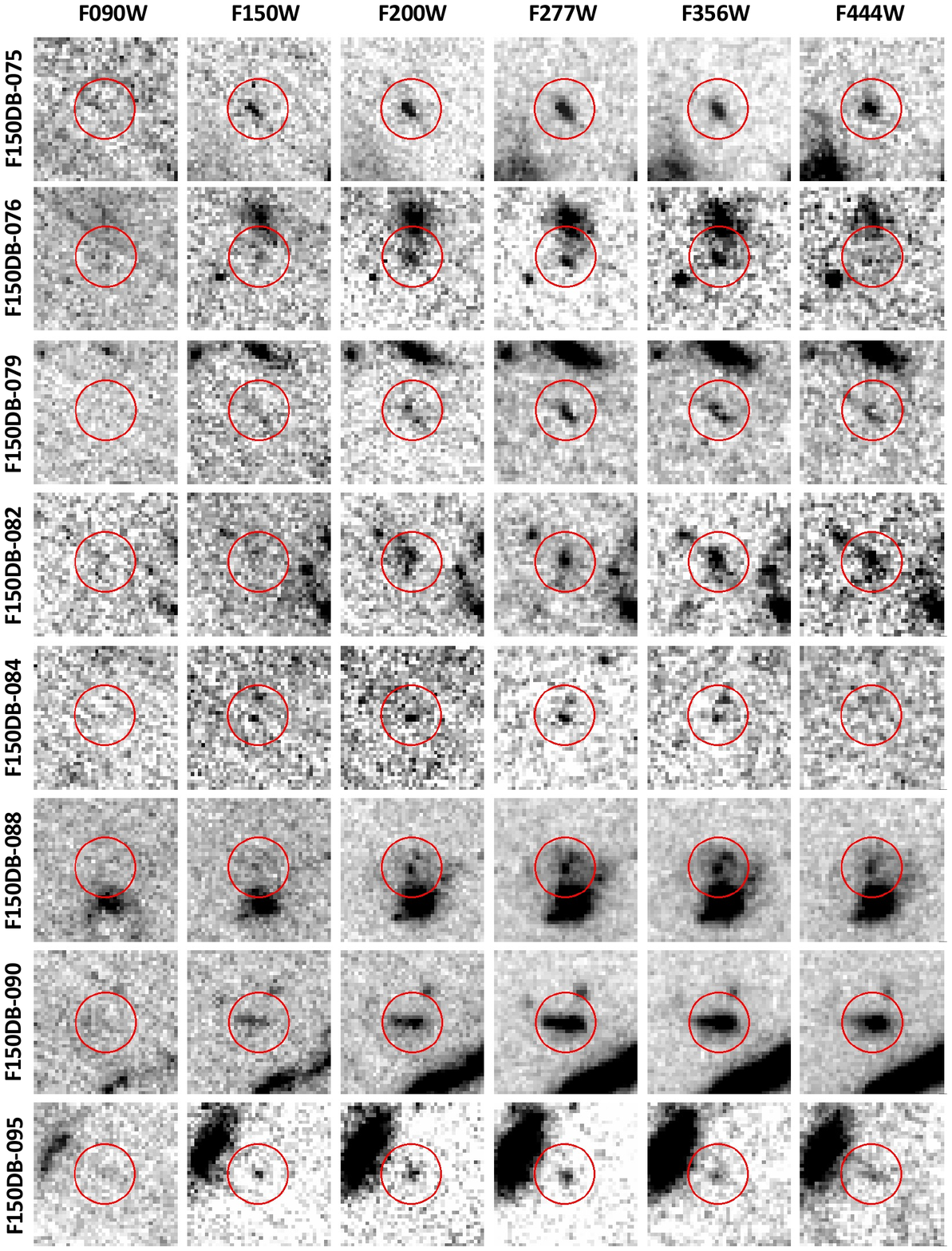}
    \caption{(cont.) 
    }
   %\label{fig:all_f150d_stamps}
\end{figure*}

\setcounter{figure}{0}
\begin{figure*}[htbp]
%\ContinuedFloat 
    \centering
    \includegraphics[height=\textheight]{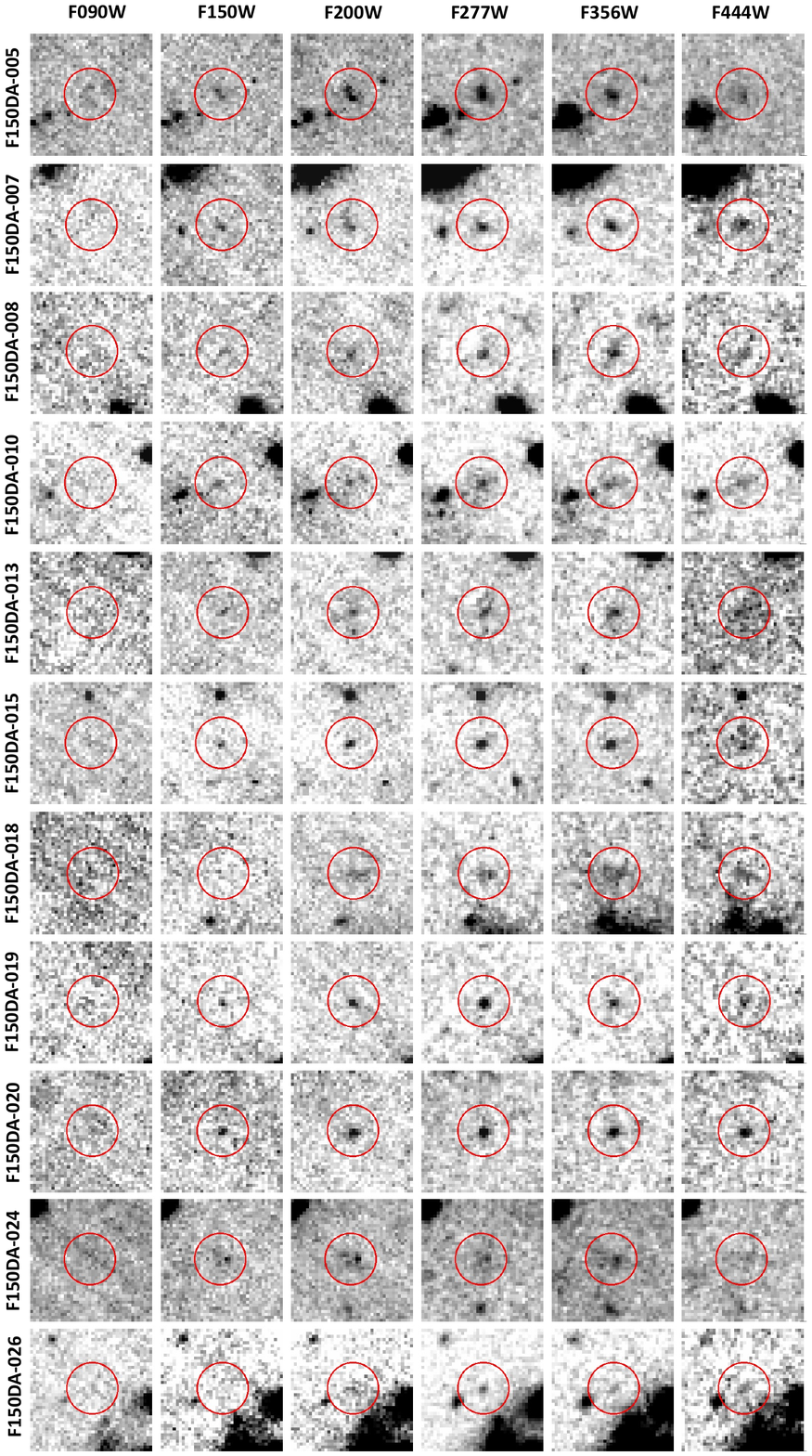}
    \caption{(cont.) 
    }
   %\label{fig:all_f150d_stamps}
\end{figure*}

\setcounter{figure}{0}
\begin{figure*}[htbp]
%\ContinuedFloat 
    \centering
    \includegraphics[height=\textheight]{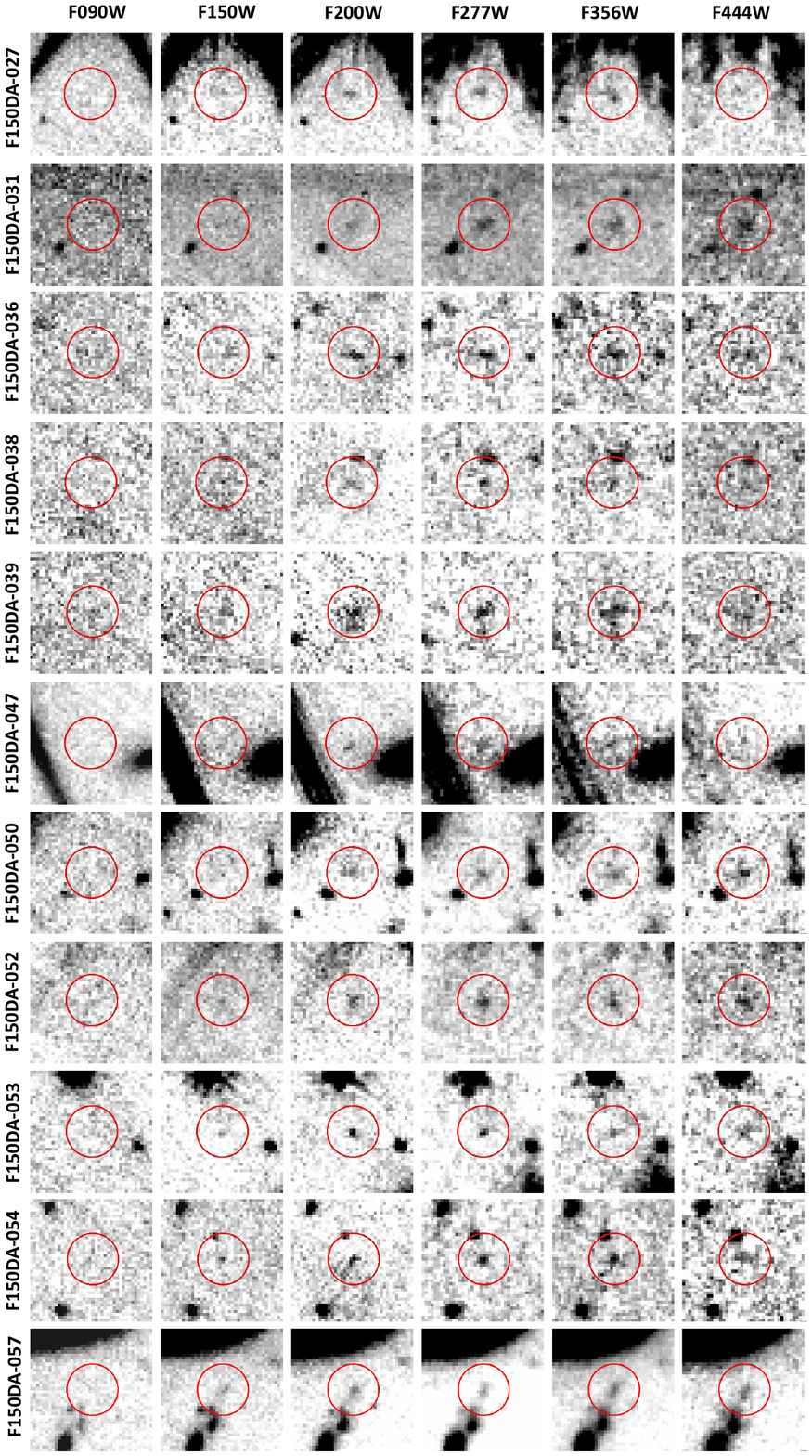}
    \caption{(cont.) 
    }
   %\label{fig:all_f150d_stamps}
\end{figure*}

\setcounter{figure}{0}
\begin{figure*}[htbp]
%\ContinuedFloat 
    \centering
    \includegraphics[height=\textheight]{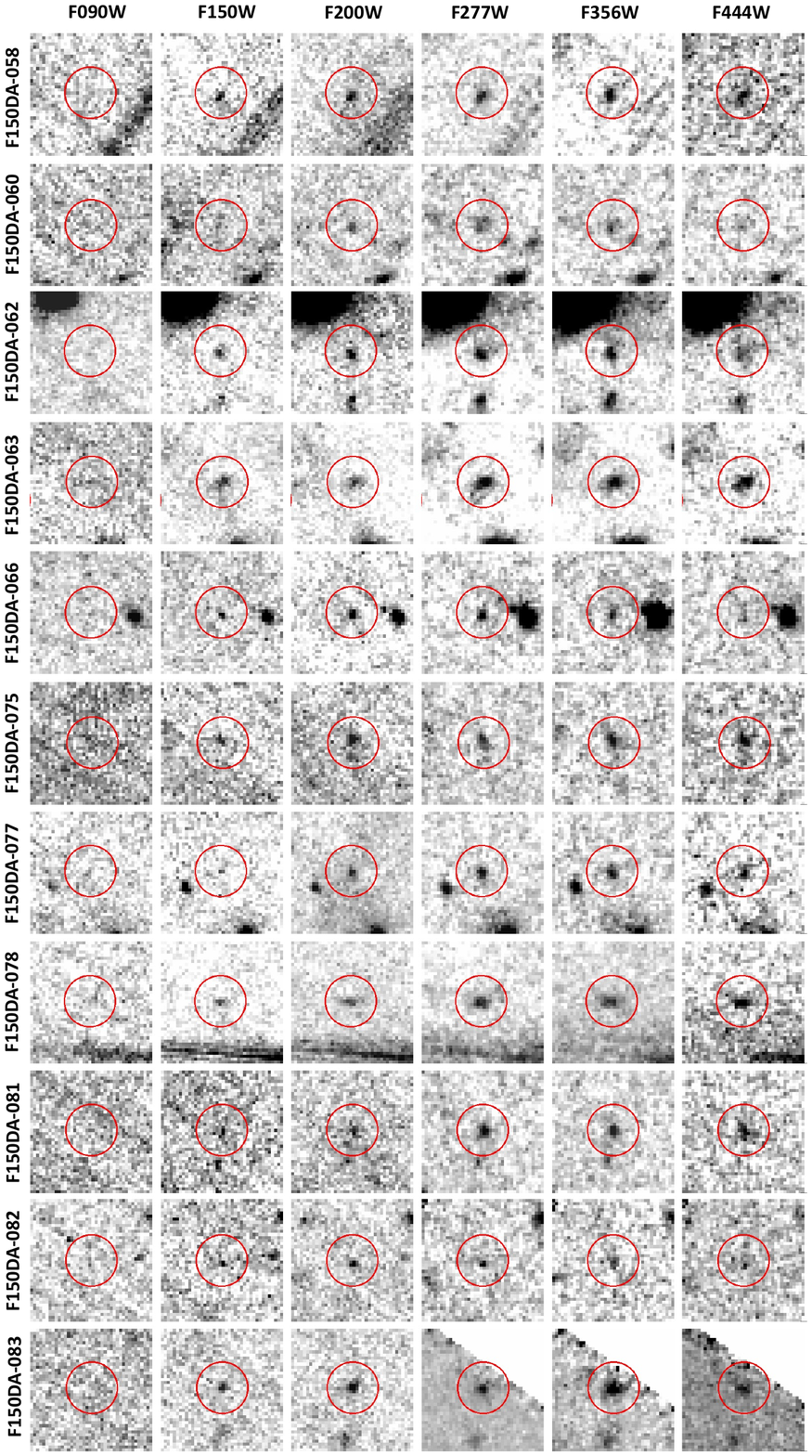}
    \caption{(cont.) 
    }
   %\label{fig:all_f150d_stamps}
\end{figure*}

\begin{figure*}[htbp]
%\ContinuedFloat
    \centering
    \includegraphics[height=\textheight]{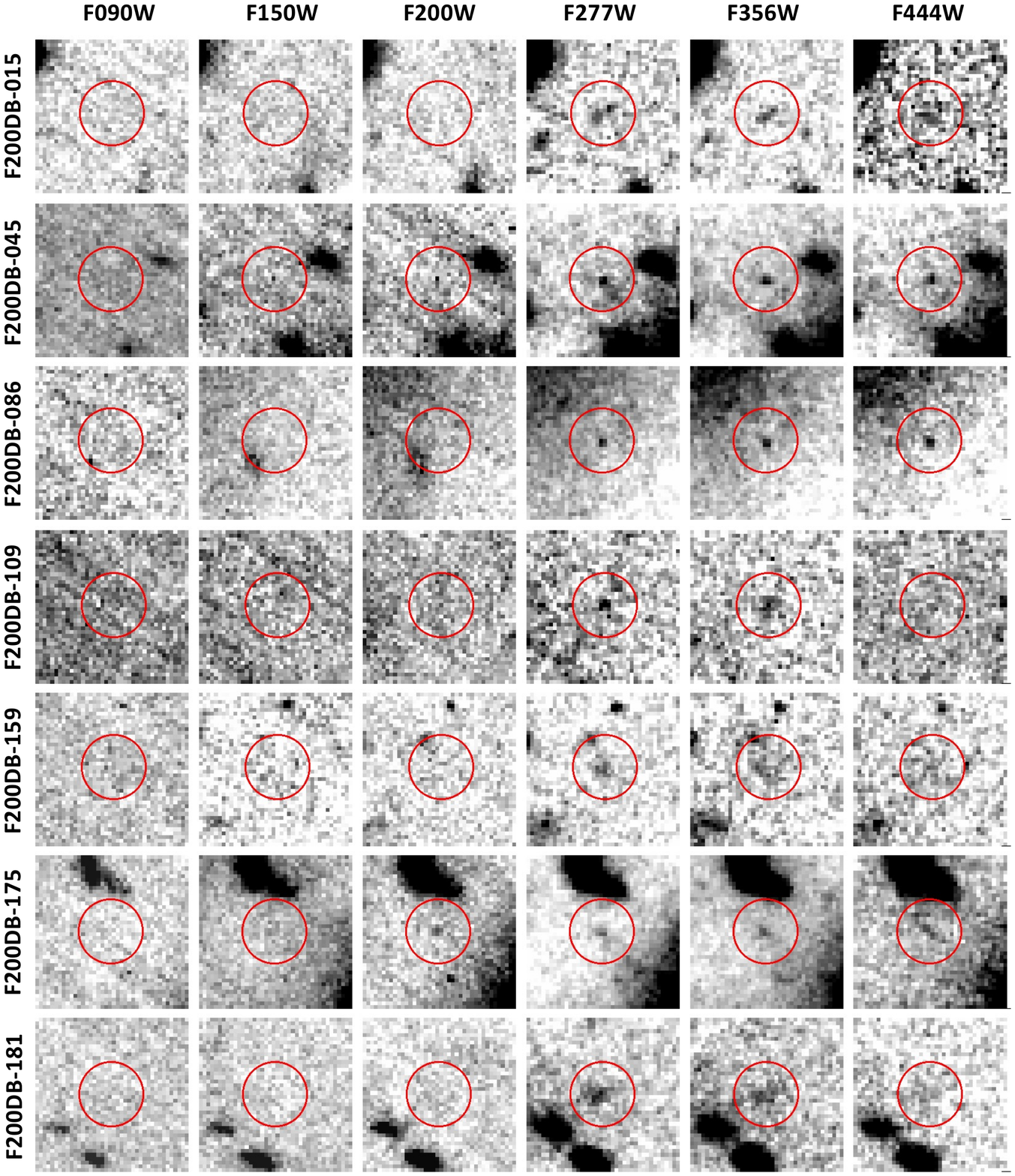}
    \caption{Similar to Figure \ref{fig:all_f150d_stamps}, but for F200W
     dropouts.
    }
   \label{fig:all_f200d_stamps}
\end{figure*}

\setcounter{figure}{1}
\begin{figure*}[htbp]
%\ContinuedFloat
    \centering
    \includegraphics[height=\textheight]{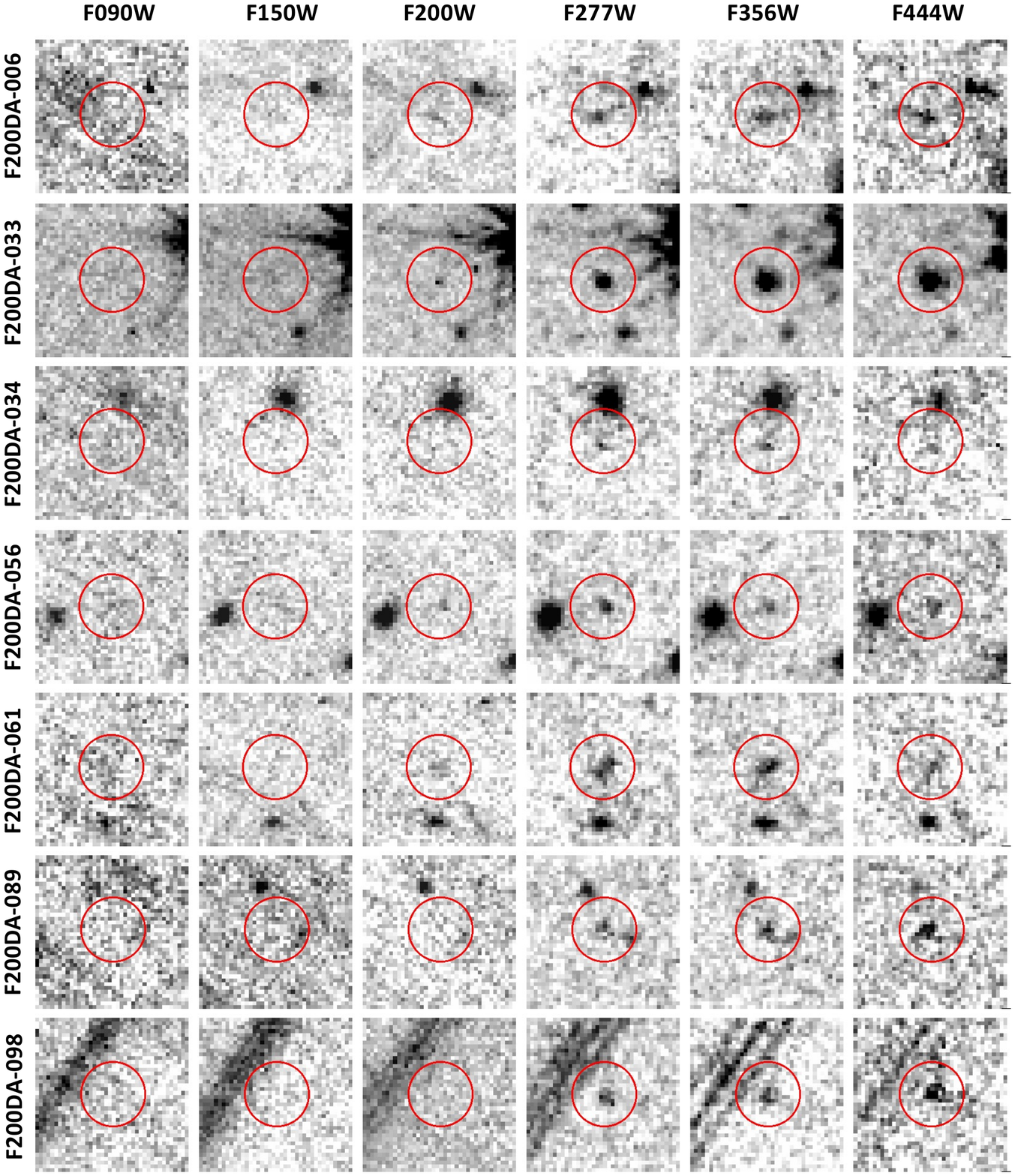}
    \caption{(cont.)
    }
   %\label{fig:all_f200d_stamps}
\end{figure*}

\begin{figure*}[htbp]
%\ContinuedFloat
    \centering
    \includegraphics[height=\textheight]{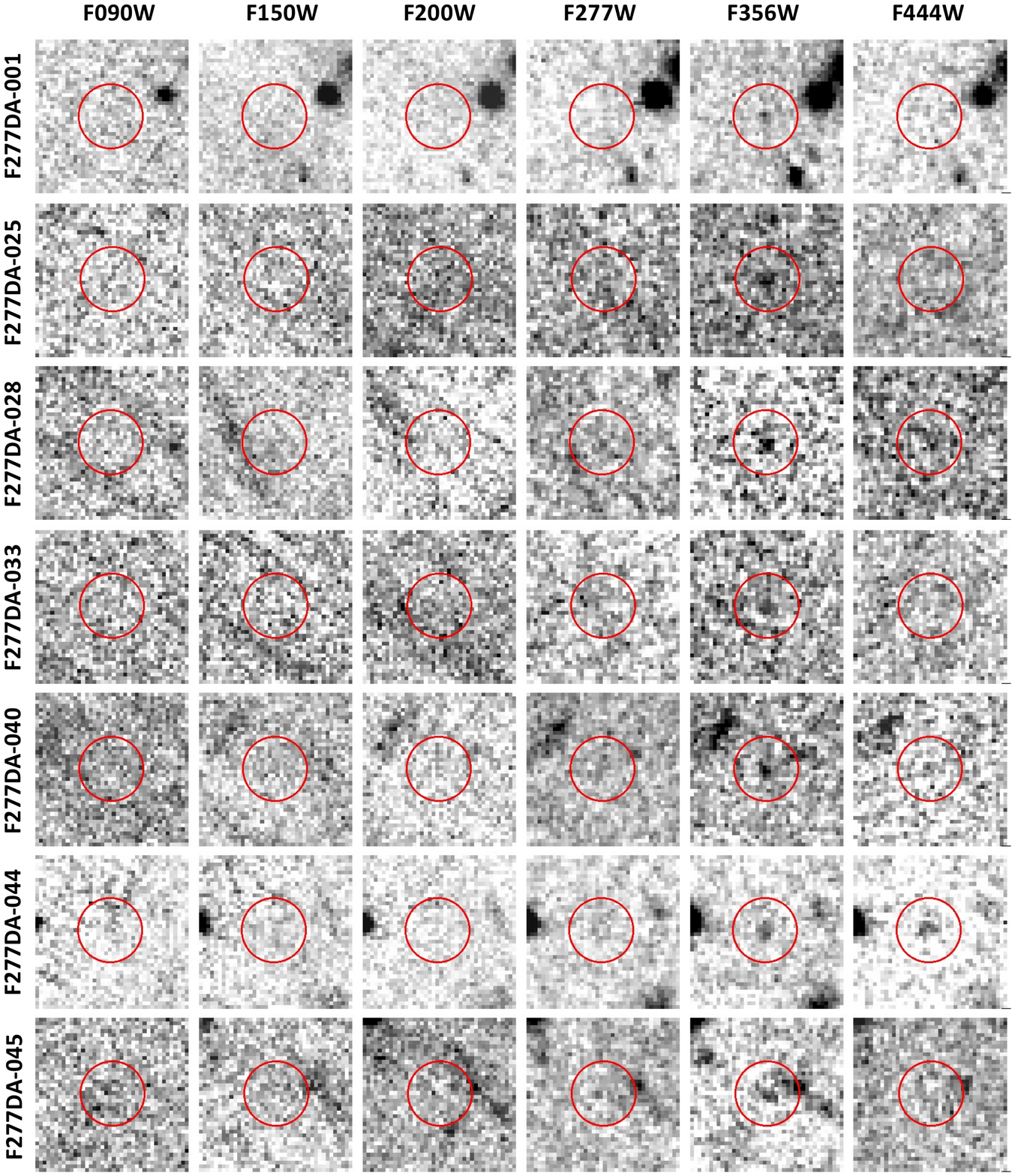}
    \caption{Similar to Figure \ref{fig:all_f150d_stamps}, but for F277W
    dropouts.
    }
   \label{fig:all_f277d_stamps}
\end{figure*}

\setcounter{figure}{2}
\begin{figure*}[htbp]
%\ContinuedFloat
    \centering
    \includegraphics[height=\textheight]{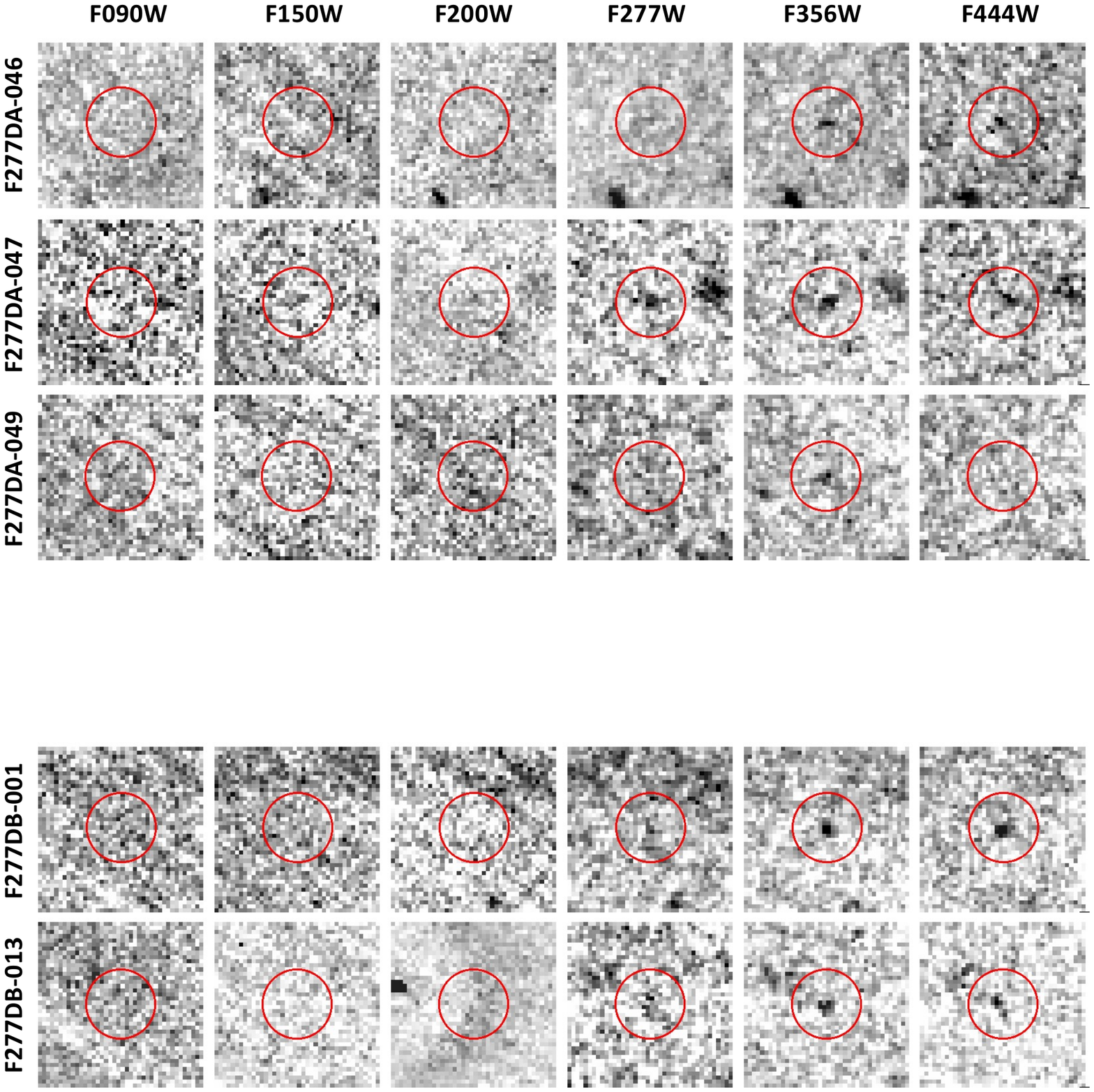}
    \caption{(cont.)
    }
  %\label{fig:all_f277d_stamps}
\end{figure*}

\setcounter{figure}{0}
\renewcommand{\thefigure}{C\arabic{figure}}

\section{Le Phare SED Fitting}

   Figure \ref{fig:sed_lephare} presents the SED fitting results for the
F150W and F200W dropouts, using Le Phare and BC03 models as described in 
Section 4.2. These are similar to those shown in Figures \ref{fig:f150d_demo}
and \ref{fig:f200d_demo}.
The ``chain of five'' are not included, as their photometry
is severely affected by blending.
The quoted $\chi^2$ are the total values (i.e., not the ``reduced $\chi^2$'').

\begin{figure*}[htbp]
%\ContinuedFloat
    \centering
    \includegraphics[height=\textheight]{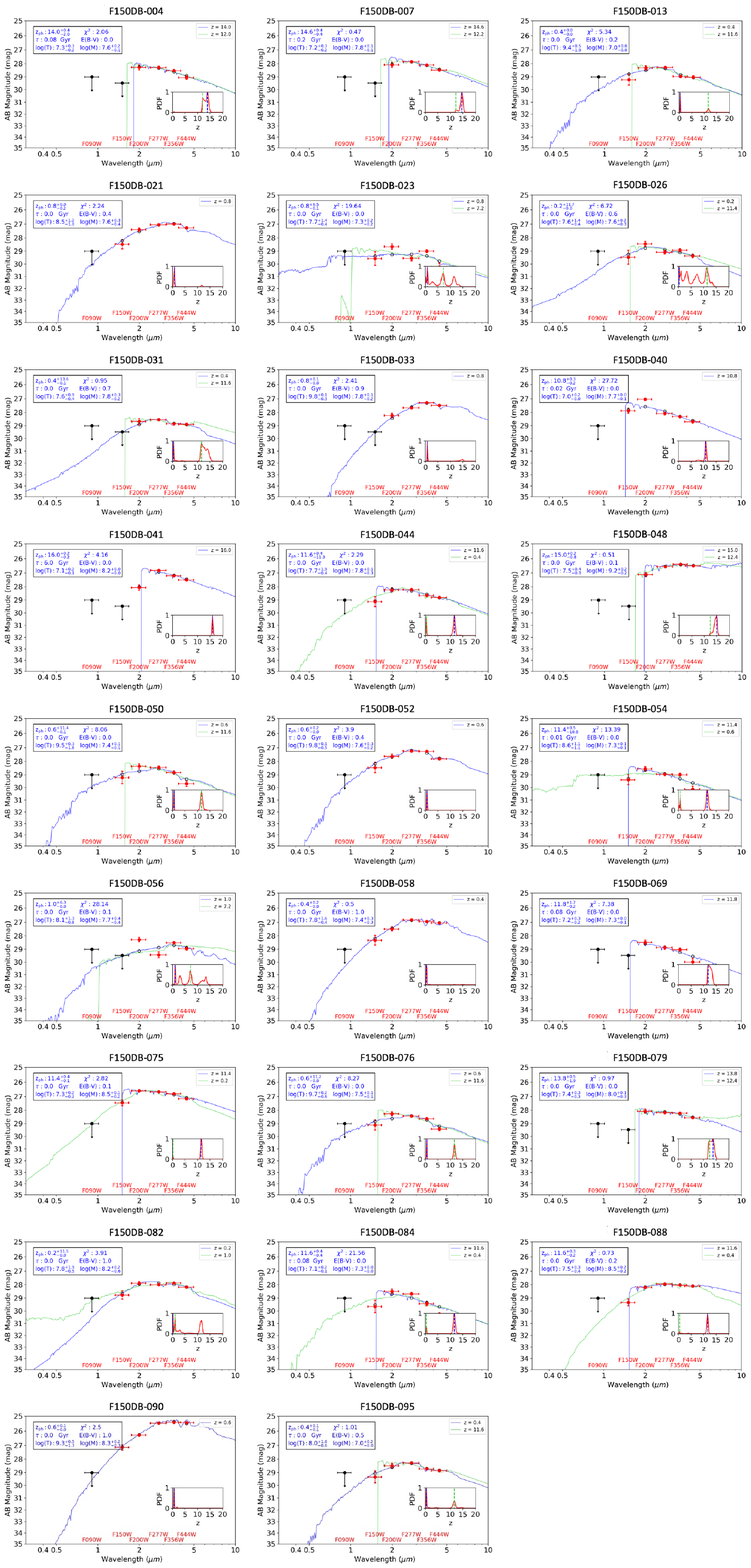}
    \caption{Le Phare SED fitting for F150W dropouts.
    }
  \label{fig:sed_lephare}
\end{figure*}

\setcounter{figure}{0}
\begin{figure*}[htbp]
%\ContinuedFloat
    \centering
    \includegraphics[height=\textheight]{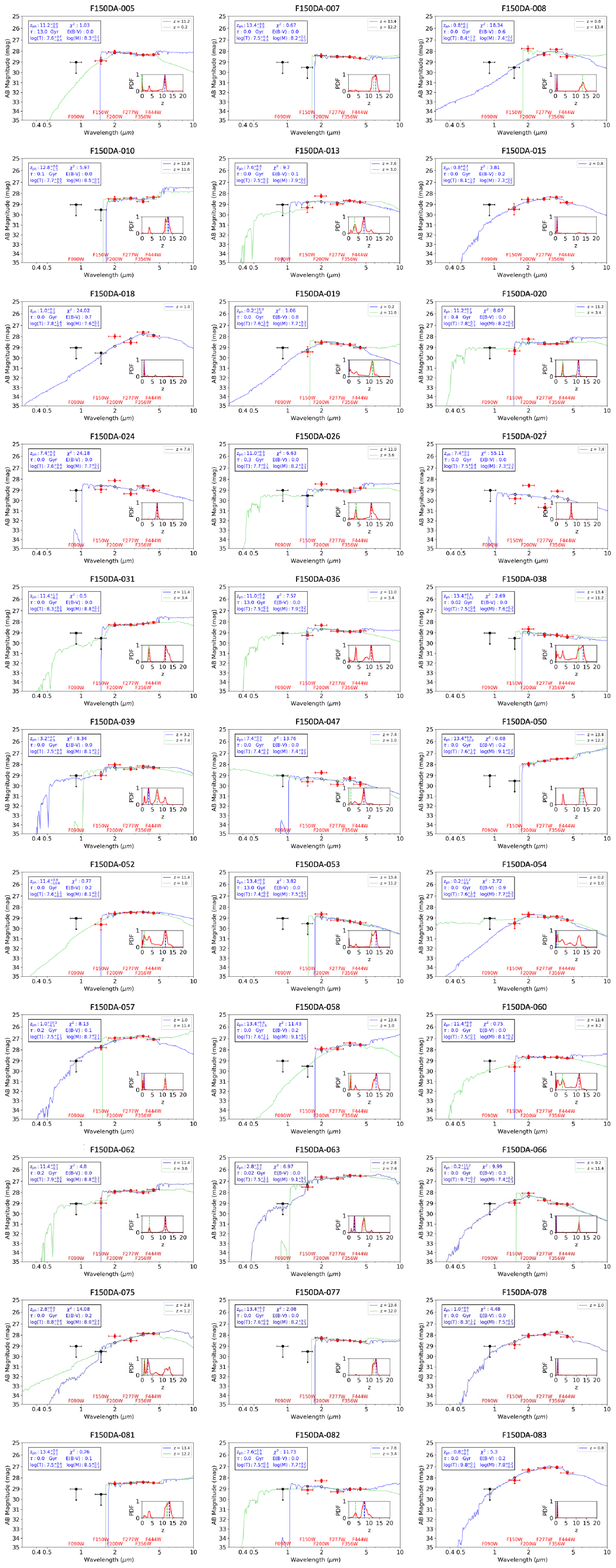}
    \caption{(cont.)
    }
  %\label{fig:sed_lephare}
\end{figure*}

\begin{figure*}[htbp]
%\ContinuedFloat
    \centering
    \includegraphics[height=\textheight]{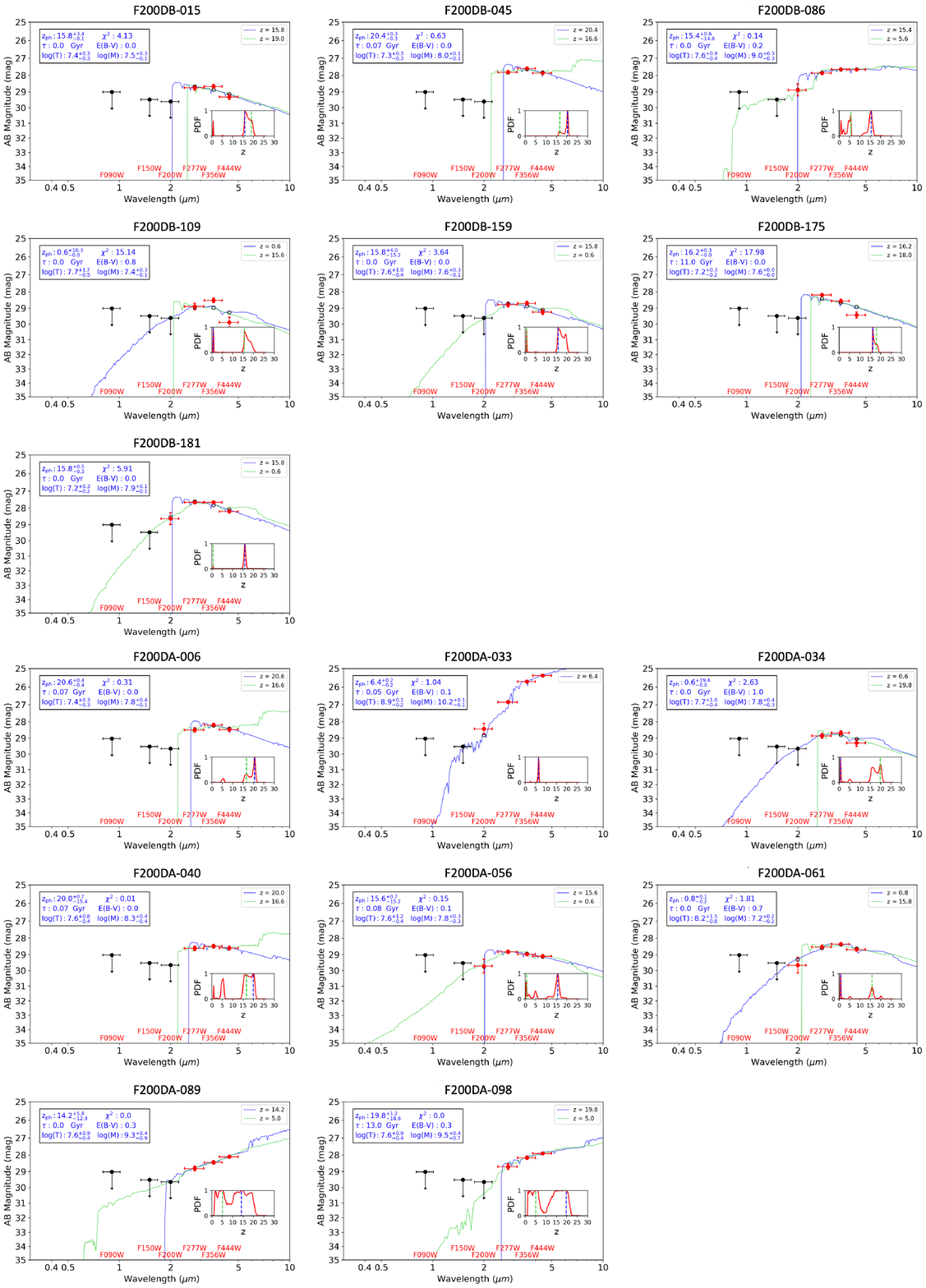}
    \caption{Le Phare SED fitting for F200W dropouts.
    }
  %\label{fig:sed_lephare}
\end{figure*}

\setcounter{figure}{0}
\renewcommand{\thefigure}{D\arabic{figure}}

\begin{figure*}[htbp]
%\ContinuedFloat
    \centering
    \includegraphics[height=\textheight]{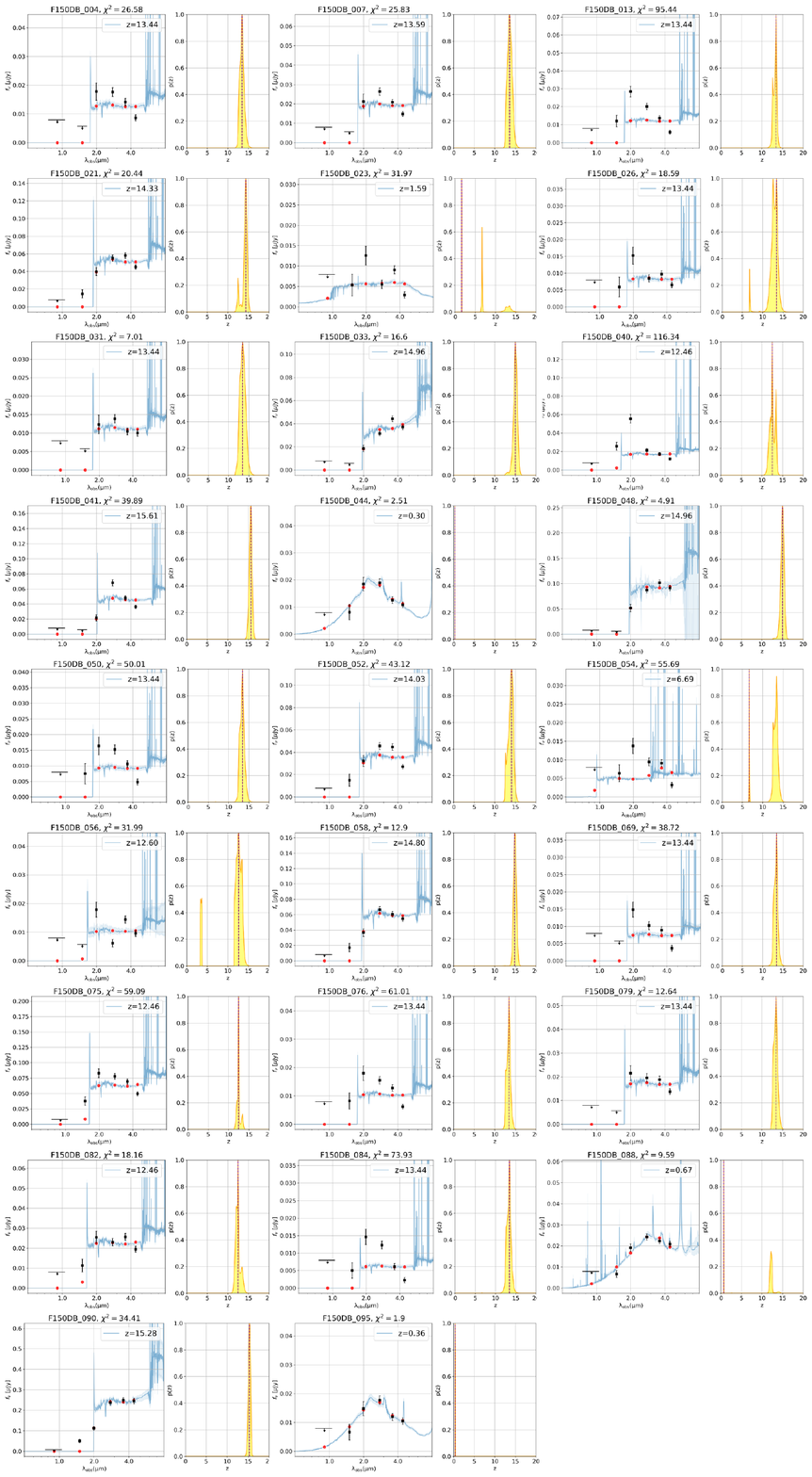}
    \caption{EAZY SED fitting for F150W dropouts.
    }
  \label{fig:sed_eazy}
\end{figure*}

\setcounter{figure}{0}
\begin{figure*}[htbp]
%\ContinuedFloat
    \centering
    \includegraphics[height=\textheight]{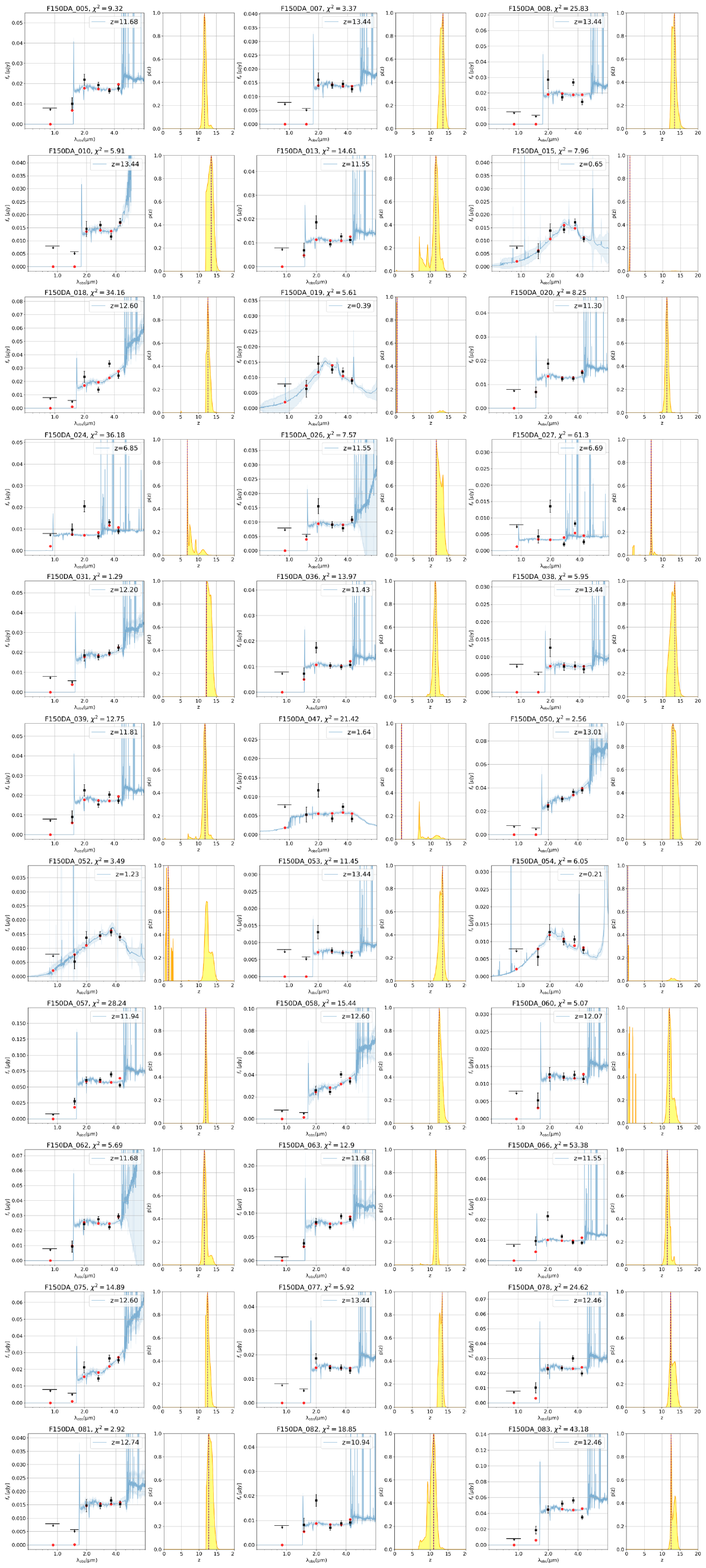}
    \caption{(cont.)
    }
 %\label{fig:sed_eazy}
\end{figure*}

\begin{figure*}[htbp]
%\ContinuedFloat
    \centering
    \includegraphics[height=\textheight]{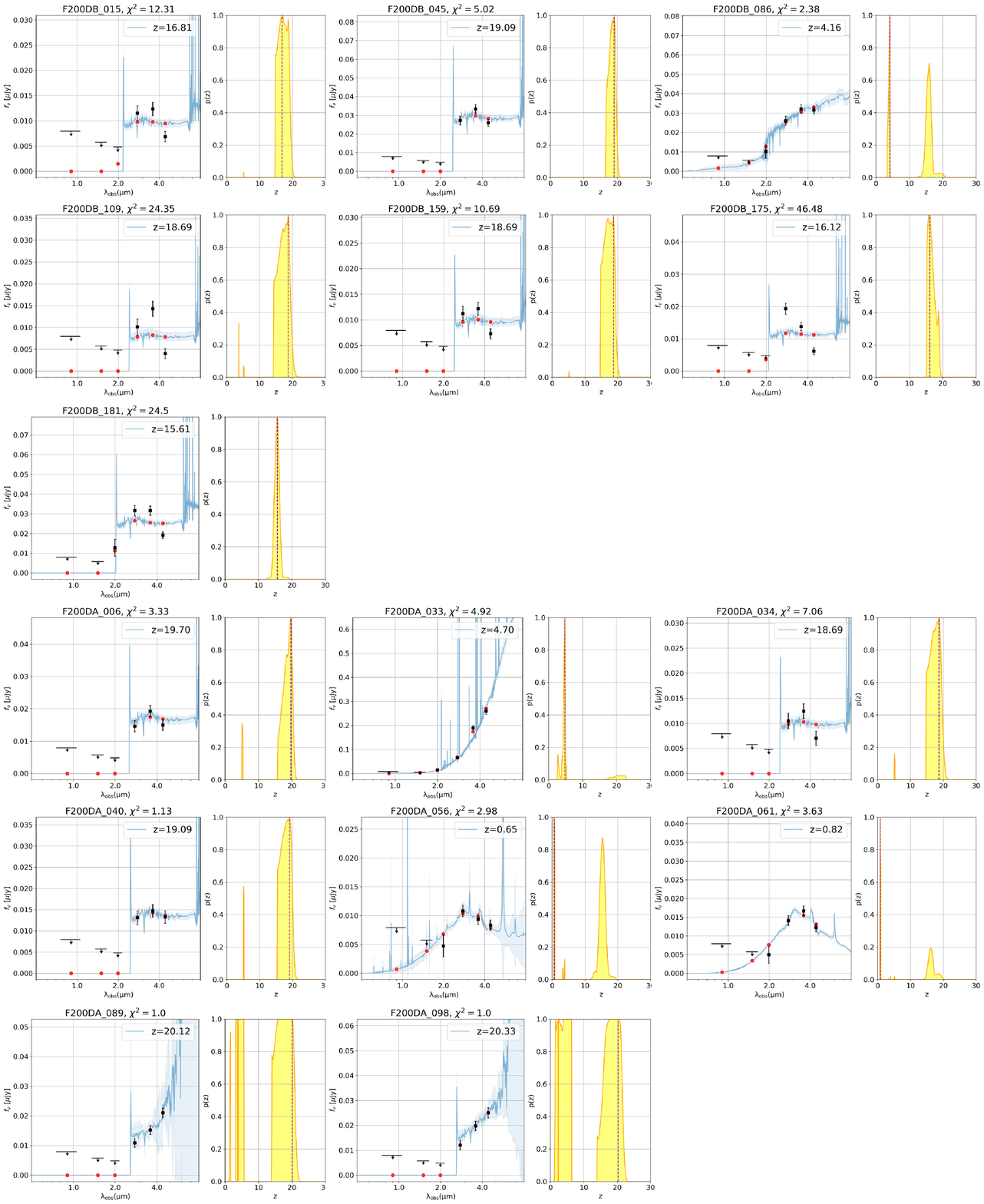}
    \caption{EAZY SED fitting for F200W dropouts.
    }
 %\label{fig:sed_eazy}
\end{figure*}

\section{EAZY SED Fitting}

   We also performed SED fitting using EAZY-py
\footnote{\url{https://github.com/gbrammer/eazy-py}}
, the latest implementation of EAZY, which incorporates templates with 
emission lines. We made a modification to the code so that flux density
upper limits can be used. By design, EAZY is for $z_{\rm ph}$ derivation but
is not for stellar population analysis. The black boxes with error bars are
the data points, and the black curves are the best-fit models. The red circles
are the synthesized flux densities based on the best-fit models. $P(z)$ is
shown next to the SED plot in each panel. The quoted $\chi^2$ are the total 
values.

\clearpage
\bibliographystyle{apj.bst}

\end{document}